\RequirePackage{fix-cm}
\documentclass[twocolumn,epjc3]{svjour3}  

\smartqed 
\RequirePackage{mathptmx}
\RequirePackage{graphicx}
\RequirePackage{rotating}
\RequirePackage{pdflscape}
\usepackage[dvipsnames]{xcolor}
\RequirePackage{lineno}
\RequirePackage{latexsym}
\RequirePackage{xspace}
\RequirePackage[numbers,sort&compress]{natbib}
\RequirePackage[colorlinks,citecolor=blue,urlcolor=blue,linkcolor=blue]{hyperref}
\usepackage{amsmath,amsfonts,bm}
\usepackage{wasysym}
\usepackage{upgreek}
\usepackage{subcaption}

\journalname{Eur. Phys. J. C}

\begin{document}
\newcommand{\DBD}{0$\nu$DBD\xspace}
\newcommand{\onbb}{$0\nu\beta\beta$\xspace}
\newcommand{\nnbb}{$2\nu\beta\beta$\xspace}
\newcommand{\thalf}{$T_{1/2}^{0\nu}$}
\newcommand{\edw}{EDELWEISS\xspace}
\newcommand{\cuore}{CUORE\xspace}
\newcommand{\cupid}{CUPID\xspace}
\newcommand{\cupido}{CUPID-0\xspace}
\newcommand{\cupidmo}{CUPID-Mo\xspace}
\newcommand{\rly}{RLY\xspace}
\newcommand{\ctsper}{counts$/($keV$\cdot$kg$\cdot$yr$)$\xspace}
\newcommand{\Qbb}{$Q_{\beta\beta}$\xspace}
\newcommand{\mbb}{$m_{\beta\beta}$\xspace}

\newcommand{\lmo}{Li$_2$MoO$_4$\xspace}
\newcommand{\enrLMO}{Li$_{2}${}$^{100}$MoO$_4$\xspace}
\newcommand{\ZS}{ZnSe\xspace}
\newcommand{\enrZS}{Zn$^{82}$Se\xspace}
\newcommand{\teo}{TeO$_2$\xspace}
\newcommand{\ntd}{NTD\xspace}

\newcommand\todo[1]{\textcolor{red}{#1}}
\newcommand\edit[1]{\textcolor{black}{#1}}
\newcommand{\tali}[1]{{\color{Magenta}{[#1] --Tali}}}

\newcommand{\Mo}{$^{100}$Mo\xspace}
\newcommand{\Te}{$^{130}$Te\xspace}
\newcommand{\Se}{$^{82}$Se\xspace}
\newcommand{\TL}{$^{208}\mathrm{Tl}$\xspace}
\newcommand{\Co}{$^{60}\mathrm{Co}$\xspace}
\newcommand{\BEGA}{$\beta/\gamma$\xspace}
\newcommand{\KF}{$^{40}\mathrm{K}$\xspace}
\newcommand{\THO}{$^{232}\mathrm{Th}$\xspace}
\newcommand{\UR}{$^{238}\mathrm{Ur}$\xspace}
\newcommand{\RA}{$^{226}\mathrm{Ra}$\xspace}
\newcommand{\PO}{$^{210}\mathrm{Po}$\xspace}
\newcommand{\PB}{$^{210}\mathrm{Pb}$\xspace}
\newcommand{\FE}{$^{55}\mathrm{Fe}$\xspace}
\newcommand{\MN}{$^{55}\mathrm{Mn}$\xspace}
\newcommand{\BI}{$^{214}\mathrm{Bi}$\xspace}
\newcommand{\PT}{$^{190}\mathrm{Pt}$\xspace}
\newcommand{\K}{$^{40}\mathrm{K}$\xspace}
\newcommand{\al}{$\alpha$\xspace}
\newcommand{\be}{$\beta$\xspace}
\newcommand{\ga}{$\gamma$\xspace}
\newcommand{\xr}{X-ray\xspace}
\newcommand{\ky}{kg$\times$yr\xspace}
\newcommand{\ckky}{counts/(keV$\times$kg$\times$yr)\xspace}
\newcommand{\cpdkk}{\un{cpd/keV/kg}\xspace}
\providecommand*{\un}[1]{\ensuremath{\mathrm{~#1}}}
\newcommand\lword[1]{\leavevmode\nobreak\hskip0pt plus\linewidth\penalty50\hskip0pt plus-\linewidth\nobreak{#1}}

\hyphenation{qua-drat-ic}
\hyphenation{con-se-quent-ly}
\hyphenation{con-stant}

\title{First demonstration of a TES based cryogenic \lmo detector for neutrinoless double beta decay search}

\author{
G.~Bratrud\thanksref{NU}\and
C.~L.~Chang\thanksref{ANL}\and
R.~Chen\thanksref{NU}\and
E.~Cudmore\thanksref{UoT}\and
E.~Figueroa-Feliciano\thanksref{NU,FNAL}\and
Z.~Hong\thanksref{UoT}\and
K.~T.~Kennard\thanksref{NU}\and
S.~Lewis\thanksref{e5,FNAL}\and
M.~Lisovenko\thanksref{ANL}\and
L.~O.~Mateo\thanksref{NU}\and
V.~Novati\thanksref{e2,e3,NU}\and
V.~Novosad\thanksref{ANL}\and
E.~Oliveri\thanksref{IJCLAB}\and
R.~Ren\thanksref{e4,NU}\and
J.~A.~Scarpaci\thanksref{IJCLAB}\and
B.~Schmidt\thanksref{e1,NU,CEA-IRFU}\and
G.~Wang\thanksref{ANL}\and
L.~Winslow\thanksref{MIT}\and
V.~G.~Yefremenko\thanksref{ANL}\and
J.~Zhang\thanksref{ANL}
\and
D.~Baxter\thanksref{FNAL,NU}\and
M.~Hollister\thanksref{FNAL}\and 
C.~James\thanksref{FNAL}\and
P.~Lukens\thanksref{FNAL}\and
D.~J.~Temples\thanksref{FNAL}
}

%Additional people: Advise on evaporation Anastasiia + anyone else, Andrea?,Claudia?
%

\thankstext{e1}{e-mail: benjamin.schmidt@cea.fr}
%\thankstext{e2}{Now at: IRFU, CEA, Universit\'{e} Paris-Saclay, F-91191 Gif-sur-Yvette, France }
\thankstext{e2}{e-mail: valentina.novati@lpsc.in2p3.fr}
\thankstext{e3}{Now at LPSC, CNRS, Universit\`e Grenoble Alpes, Grenoble, France}
\thankstext{e4}{Now at  University of Toronto, Toronto, Canada}
\thankstext{e5}{Now at Wellesley College, Wellesley, MA 02481, USA }
\institute{
Northwestern University, 633 Clark St, Evanston, IL 60208, USA \label{NU}
\and 
Argonne National Laboratory, 9700 S Cass Ave, Lemont, IL 60439, USA \label{ANL}
\and
Department of Physics, University of Toronto, 27 King’s College Cir,15
Toronto, M5R 0A3, ON, Canada \label{UoT}
\and 
Universit\'e Paris-Saclay, CNRS/IN2P3, IJCLab, 91405 Orsay, France \label{IJCLAB}
\and
Massachusetts Institute of Technology, Cambridge, MA 02139, USA \label{MIT} 
\and
IRFU, CEA, Universit\'{e} Paris-Saclay, F-91191 Gif-sur-Yvette, France  \label{CEA-IRFU} 
\and
Fermi National Accelerator Laboratory, Batavia, Il, USA\label{FNAL}
}

\date{Received: date / Accepted: date}

\maketitle

\begin{abstract}
 Cryogenic calorimetric experiments to search for neutrinoless double-beta decay (\onbb) are highly competitive, scalable and versatile in isotope.  The largest planned detector array, CUPID, is comprised of about 1500 individual \enrLMO detector modules with a further scale up envisioned for a follow up experiment (CUPID-1T). 
In this article, we present a novel detector concept targeting this second stage with a low impedance TES based readout for the \lmo absorber that is easily mass-produced and lends itself to a multiplexed readout. We present the detector design and results from a first prototype detector operated at the NEXUS shallow underground facility at Fermilab. 
The detector is a 2-cm-side cube with 21\,g mass that is strongly thermally coupled to its readout chip to allow rise-times of $\sim$0.5\,ms. This design is more than one order of magnitude faster than present NTD based detectors and is hence expected to effectively mitigate backgrounds generated through the pile-up of two independent two neutrino decay events coinciding close in time. 
Together with a baseline resolution of 1.95\,keV (FWHM) these performance parameters extrapolate to a background index from pile-up as low as $5\cdot 10^{-6}\,$counts/keV/kg/yr in CUPID size crystals. The detector was calibrated up to the MeV region showing sufficient dynamic range for \onbb searches. 
In combination with a SuperCDMS HVeV detector this setup also allowed us to perform a precision measurement of the scintillation time constants of \lmo, which showed a primary component with a fast O(20~$\mu$s) time scale. 

\end{abstract}


\section{Introduction}
\label{sec:intro}
Neutrinoless double beta decay (\onbb) is a hypothetical nuclear transition $A \Longrightarrow (A-2) + 2 e^{-}$ that shares the full decay-energy between the daughter nucleus and the two electrons, without emission of electron anti-neutrinos. 
It violates lepton number by two units and would constitute clear evidence of physics beyond the standard model. 
The observation of this decay was first proposed by Furry in 1939~\cite{Furry:1939}, two years after Ettore Majorana pointed out that the neutrino, being a neutral lepton, could be described by a real valued wave function as solution of a Hermitian form of the Dirac equation. 
This in turn implies that these solutions are identical to the solutions for their anti-particles \cite{Majorana:1937}. 
As a massive particle this Majorana neutrino would interact both as a left-handed particle and a right-handed anti-particle. 
Consequently it would allow for a double beta decay process without emission of neutrinos in addition to the standard-model-allowed double beta decay process with emission of two neutrinos \cite{GoeppertMayer:1935}. 
Beyond this so called mass mechanism several other standard model extensions provide lepton number violating mechanisms \cite{Vergados:2016,DellOro:2016tmg,Bilenky:2015,Deppisch:2012,Rodejohann:2012} that can be tested through neutrinoless double beta decay searches. \onbb is hence a crucial probe that complements direct \cite{Aker:2022, Formaggio:2021} and indirect neutrino mass measurements \cite{PDG:2022}. It is uniquely placed to answer the question if there is a Majorana neutrino mass and if neutrinos may have contributed to the generation of the matter-antimatter asymmetry in the early Universe \cite{Drewes:2016}. 

Cryogenic calorimeters are among the most flexible technologies to search for \onbb \cite{Fiorini:1984} allowing the study of different isotopes with the source isotopes embedded in the detector. 
Currently cryogenic calorimeters provide leading results for \Te, \Se and \Mo \cite{Adams:2022, Azzolini:2022, Augier:2022}. 
CUORE (Cryogenic Underground Observatory for Rare Events) has demonstrated the efficient scalability of the complex cryogenic infrastructure to deploy a tonne-scale of isotope of interest \cite{Alduino:2018,ADAMS:2022a} and to operate it with high efficiency and up-time \cite{Adams:2020, Adams:2022}. 
The CUPID-Mo \cite{Armengaud:2020} and CUPID-0 \cite{Azzolini:2018tum} experiments have provided a robust demonstration of the reduction of previously leading backgrounds from $\alpha$-decays \cite{Azzolini:2019nmi,Augier:2023} with the use of scintillating bolometers with a dual readout of both the heat and light signals. 
The deployment of this technology in the CUORE Upgrade with Particle Identification (CUPID) \cite{CUPIDInterestGroup:2019inu} experiment is expected to provide two orders of magnitude in background improvement compared to CUORE. 
However, due to the comparatively fast half-life of the \Mo \nnbb decay \cite{Augier:2023a,Arnold:2019}, paired with the slow rise-times of O(10\,ms) of the Neutron Transmutation Doped germanium thermistor \cite{Haller:1994} readout of the phonon signal, this pile-up background of random \nnbb coincidences has to be considered. 
This background has been observed in the CUPID-Mo demonstrator \cite{Augier:2023} and is projected to constitute the dominant background for CUPID: 
a random coincidence of two pulses, happening very close in time is reconstructed as a single event with increased pulse height. 
The reconstructed events thus extend into the Region Of Interest (ROI) for the \onbb signal and beyond.
The use of light detectors with boosted signal to noise \cite{Novati:2019} ratio will allow CUPID to reach a background index below its goal of $5\cdot10^{-5}$\,counts/keV/kg/yr, utilizing the faster rise-time of the germanium light detectors \cite{Chernyak:2017} as opposed to the use of the slower signal of the \lmo detectors \cite{CUPIDInterestGroup:2019inu,Chernyak:2012,Armatol:2021}.
However, looking towards a future experiment beyond CUPID with a tonne or more of \Mo, pile-up requirements below the $10^{-5}$\,counts/keV/kg/yr level are required.
This could be achieved with smaller \lmo crystals but at the expense of the \onbb containment efficiency and a further increasing channel count conflicting with very stringent cryogenic and radiopurity requirements. 
Alternatively, other isotopes could be studied with cryogenic bolometers. However, \Mo based searches have several advantages. 
The high Q-value of 3034\,keV \cite{Rahaman:2008} lies above the endpoint of the intense $\gamma$ lines in the natural U- and Th-chains allowing for a background free search. \Mo has a reasonable isotopic abundance of 9.7\%, is compatible with existing technologies for isotope enrichment \cite{CUPIDInterestGroup:2019inu}, and has excellent properties in terms of the radiopure Czochralski and Bridgman growth of Mo based crystals~\cite{Armengaud:2017}. 
Last but not least, the \onbb transition in \Mo provides a favorable product of phase space factor and nuclear transition matrix element, which results in a very good sensitivity on the effective Majorana neutrino mass per mol of isotope \cite{Agostini:2023}.
All these considerations combined have sparked a significant interest in technologies that can help resolve this pile-up background through faster sensors. 
Low-impedance sensors coupled to SQUID amplifiers with intrinsically high bandwidth are hence a desirable alternative to NTD sensors. In addition, this would also open up the possibility of employing a multiplexed readout system with reduced thermal requirements for the cryogenic system. This approach is being pursued with metallic magnetic calorimeter (MMC) sensors for both the main crystal as well as the wafer acting as light detector in the AMoRE experiment \cite{Kim:2017, Kim:2022,Kim:2023} and with Kinetic Inductance Detectors KIDs as used in \cite{Casali:2019}. However, while significant progress in detector performance both in terms of energy resolution and radio-purity has been shown within the AMoRE program, these technologies have yet to demonstrate equivalent performance to the NTD based detectors operated by the CUPID-Mo and CUPID collaborations \cite{Armengaud:2020,Armengaud:2021,Augier:2022,Alfonso:2023,Alfonso:2022,Armatol:2021b}. 
Another alternative is being investigated in the form of IrPt bi-layer transition edge sensors deposited directly onto a silicon wafer as potential light detectors for CUPID \cite{Singh:2023}. 
Here, we will focus on the investigation of a new AlMn TES chip design that has recently been developed for the Ricochet experiment \cite{Chen:2023}. It boasts a massive scalability of fabrication, where O(1000) TES chips can be produced on a single wafer and they can be coupled to different targets as for example silicon \cite{Doug:2023}, germanium, 
or \lmo as presented in this article. In Sec.~\ref{sec:experiment} we will describe the preparation of the detector module out of a 2\,cm \lmo cube, its thermal and electrical connections as well as its operation in the Northwestern EXperimental Underground Site (NEXUS)
cryogenic facility. In Sec.~\ref{sec:performance}, we will detail the detector response in time and energy and discuss the dynamic range of the device. We assess the crystal performance in terms of its scintillation time constants providing the first measurement of these properties for a crystal from the crystal growth program of Radiation Monitoring Devices, Inc. (RMD) for CUPID. Finally, we will discuss the further development of this technology in Sec.~\ref{sec:outlook}. 

\section{Experimental setup}
\label{sec:experiment}

In the TES development program for Ricochet, target crystals ranging from $1 \times 1 \times 0.4$\,cm$^3$ to $2 \times 2 \times 2$\,cm$^3$ have been utilized. 
They are prepared and wire-bonded at Northwestern University in Evanston and then transported and measured in the NEXUS cryogenic test facility at Fermilab. 
In the following, we will detail the crystal preparation (sec.~\ref{sub:crystal}), the design of the assembly with respect to electrical and thermal properties (sec.~\ref{sub:assembly}), the assembly of the \lmo crystal and the SuperCDMS HVeV detector used as corresponding light detector (sec.~\ref{sub:LightDetector}) and  finally the  cryogenic facility with its shielding and readout (sec.~\ref{sub:NEXUS}).

%-----------------------------------------------
\subsection{Crystal preparation}
\label{sub:crystal}

The TES based readout scheme developed for the Ricochet experiment uses a doped Al(Mn) bi-layer TES with 2000 ppm and 2500 ppm of Mn dopant within the respective layers. It is fabricated as a large array on a silicon carrier chip \cite{Doug:2023, Chen:2023}. 
This allows for scalable mass-production and simplified preparation with less fabrication steps for the main absorber, which only needs to be connected to the readout chip. 
While this connection can be done with glue as described in \cite{Angloher:2023} similar to the connection of NTD's, a stronger more reproducible heat conductance can be achieved with a metal film evaporated onto the target.
\begin{figure}[htbp]
\centering
\includegraphics[width=0.48\textwidth]{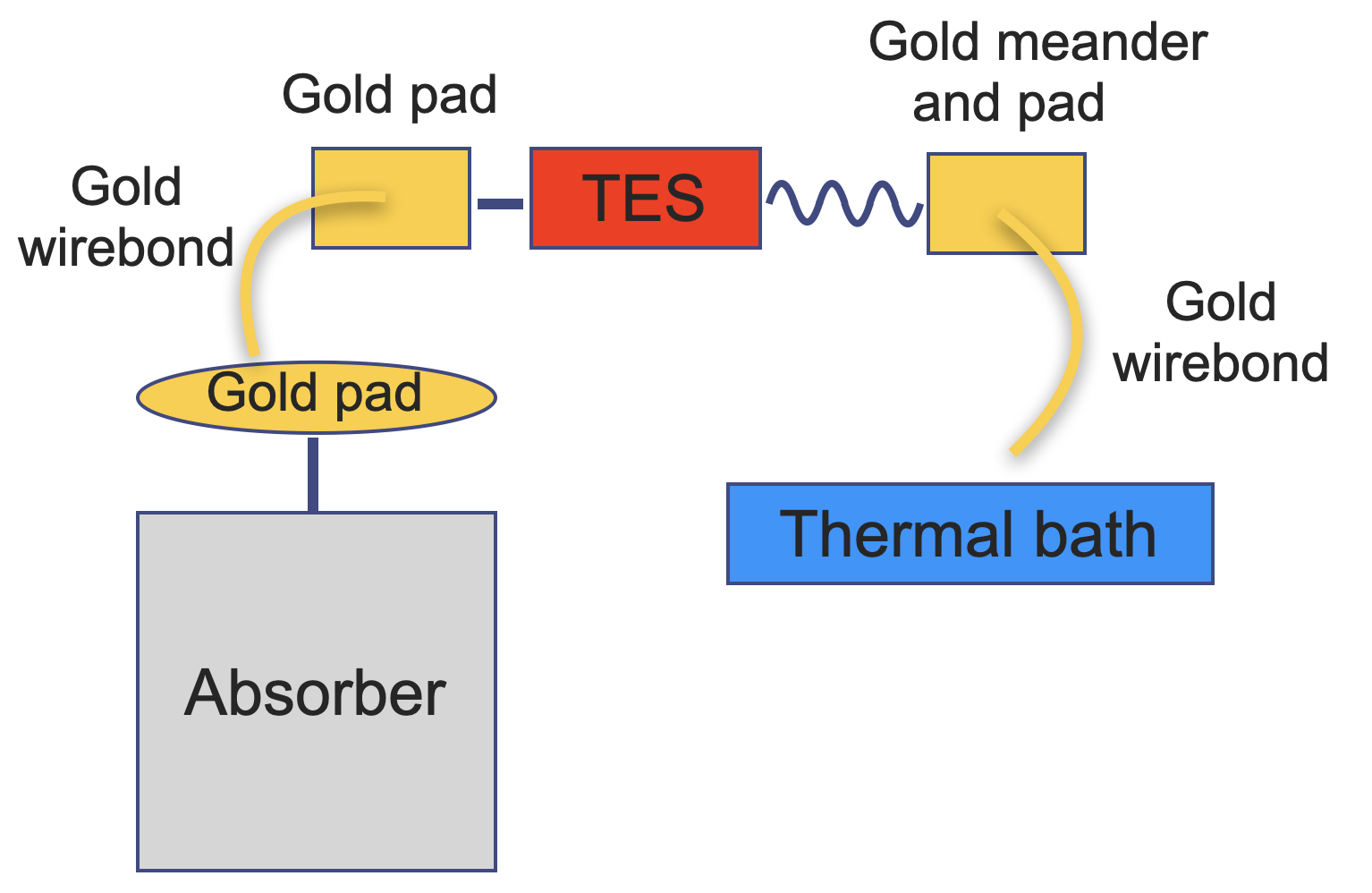}

\caption{ Simplified thermal scheme, showing the design idea for the coupling of the TES sensor to the \lmo absorber. Energy from a particle interaction is converted into phonons in the absorber, coupled to the electron system in the circular gold pad and then transported via gold wirebonds through the TES sensor to the bath. The meander between the TES and second gold pad represents a thermal impedance that is designed to ensure that the TES and absorber stay in thermal equilibrium. }
\label{fig:HeatModel}
\end{figure}

The thermal circuit has been described in detail in \cite{Chen:2023, Chen:2023b}. It was designed recently for Ricochet, but we note independent detector developments in AMoRE \cite{Kim:2017} and of the remoTES \cite{Angloher:2023}, which follow a similar design principle.  A simplified block diagram of the heat flow is shown in Fig.~\ref{fig:HeatModel}:  
An initial energy deposit from a particle interaction in a target crystal (Absorber) will quickly thermalize and form a thermal phonon bath in the crystal. In order to transport this energy to the TES sensor, a fraction of one surface is covered with a thin $\sim400\,\mathrm{nm}$ gold pad. The volume of this film is a compromise as it needs to be sufficiently large to surpass the very weak electron-phonon coupling at $\sim$20\,mK, while still adding a limited heat capacity to the main absorber. Taking values for the specific heat of pure gold from~\cite{Gopal:2012} we expect the gold film to contribute less than half of the total heat capacity for our prototype.
Once the energy of the interaction is  coupled to the electron system of the gold it can be effectively transported to the TES on its carrier chip through gold wirebonds. 
To guarantee that the TES is well thermalized to the absorber and to optimize the response and bandwidth of the detector, a thermal impedance has been carefully designed and added on the outgoing connection from the TES to the thermal bath \cite{Chen:2023, Chen:2023b}.
It is made through a narrow gold trace deposited onto the chip (also called meander) and ultimately connected to the bath via a second gold wirebond.   

For the present test, we use a cubic \lmo crystal with 2\,cm edge length (21\,g mass) grown with the Czochralski method by RMD. 
It was grown as part of an early batch within the crystal vendor validation campaign for CUPID. Since \lmo is slightly hygroscopic, the crystal was kept in a dry atmosphere of N$_2$ boil-off in a purge box until deposition of the gold, assembly and transport to NEXUS.
The gold film deposition was performed at NuFab in an e-beam evaporator with a shadow mask as a circle with 1\,cm diameter and 400\,nm thickness:  First, the crystal was exposed to a 10\,minute argon plasma cleaning in a SAMCO PC-300 plasma cleaner, to remove any residual surface contamination. We used a chamber pressure of 11\,Pa and 100\,W power. Second, the crystal was transferred into the AJA e-beam evaporator at NuFab to deposit a 3\,nm titanium adhesion layer and the 400\,nm thick gold layer. To avoid a thermal shock after deposition, the crystal was left in the deposition chamber for an additional 5\,minutes before gently flushing the chamber with N$_2$.

\begin{figure*}[ht]
\centering
\includegraphics[height=0.325\textheight]{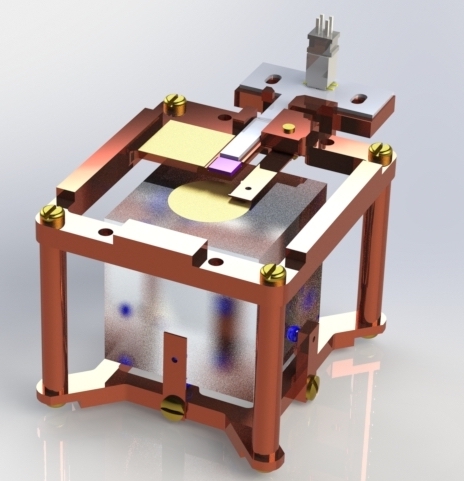}
\includegraphics[height=0.325\textheight]{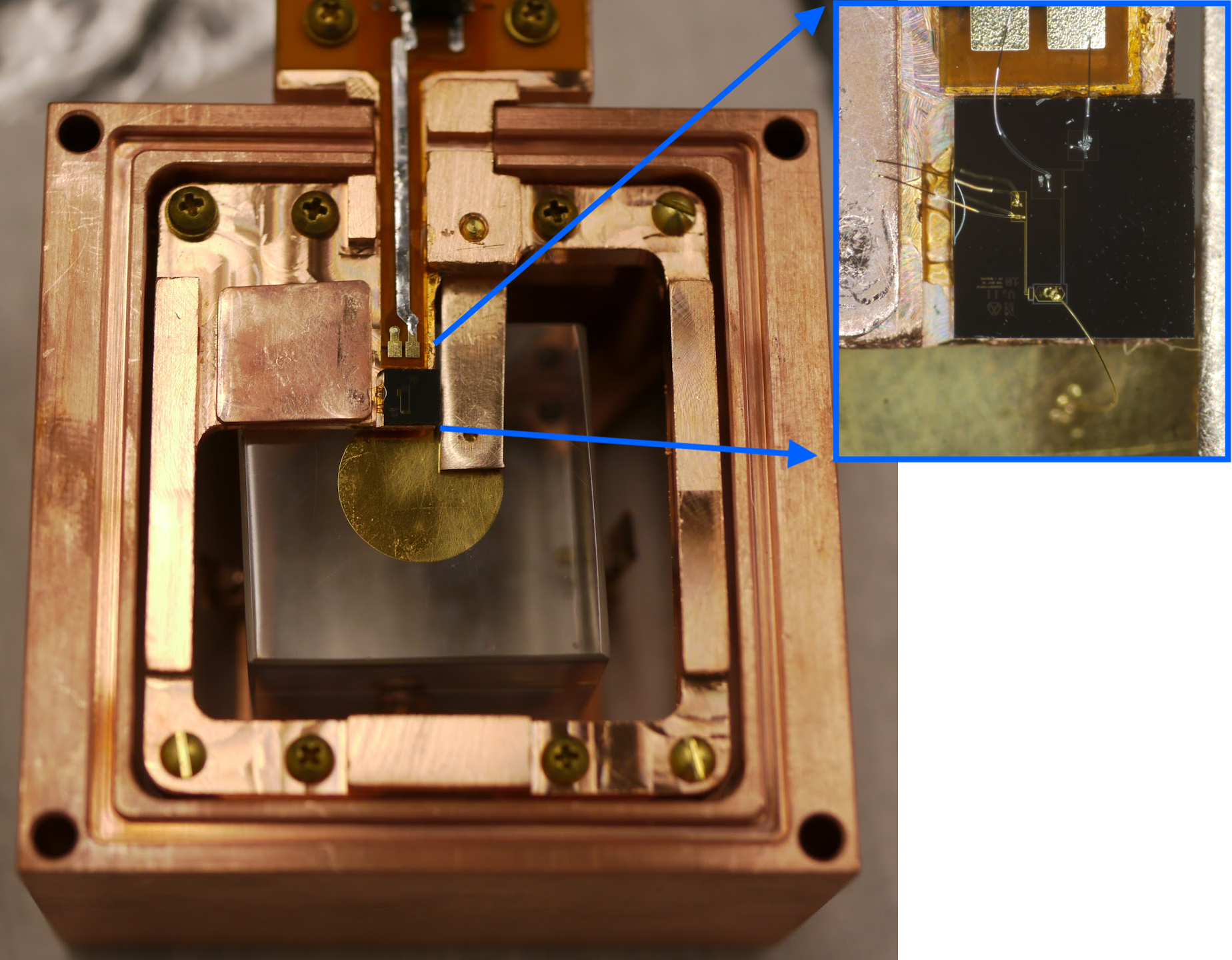}
\caption{\lmo detector module. Left: False color CAD rendering of the inner part of the \lmo detector module. The main elements of the the thermal circuit are the \lmo absorber (transparent) with its circular gold deposition on top. It is connected to the TES chip (purple) through a gold wirebond  and heat sunk to a di-electrically decoupled thermal bath (golden rectangle on the top copper piece). A custom flex PCB (gray) with tinned traces provides a superconducting connection to the SQUID readout. Right: Image of the detector after assembly including a zoom to the TES chip and its wirebonds for the thermal and electrical connection.}
\label{fig:module}
\end{figure*}

%-----------------------------------------------
\subsection{Design of the detector assembly}
\label{sub:assembly}

The detector assembly has been optimized for the prototyping and testing of novel detectors of different size and mass without strict protocols for extremely low radio-purity levels. 
A rendering together with the final assembly of the \lmo detector is shown in Fig.~\ref{fig:module}. The holder makes use of 8 sapphire spheres to thermally isolate the detector from the copper structure connected to the mixing chamber stage of the cryostat \cite{Doug:2022}.  Their size was selected in the range of 1\,mm - 3\,mm diameter to compensate for irregularities in the crystal dimension. To minimize stress from thermal contraction and a potentially connected excess of low  energy events  \cite{Adari:2022, Astrom:2006}, both the top and side spheres are held in place by flexible phosphor bronze springs designed to retain the crystal with a force of about 3\,N each.

The assembly is made up of three pieces of oxygen free copper (C101, 99.99\% purity). The top plate carries the TES readout chip (visualized in the rendering in Fig.~\ref{fig:module} in pink), its readout via a custom designed flexible polyimide PCB (gray) and a di-electrically decoupled copper plate (gold). This plate allows us to complete the thermal circuit from the sensor to the bath, while maintaining the sensor ground separate from the ground of the cryostat. In the final assembly (Fig.~\ref{fig:module} right) this plate is cut from oxygen free copper and electrically decoupled through a very thin film of 1266 Stycast with an admixture of  20\,$\mathrm{\upmu m}$ silica beads to maintain a minimal thickness.
The design allows for a consistent minimal distance of the thermal connection from the absorber's gold pad to the TES sensor of $\sim$2\,mm through a gold wire-bond and a similar connection from the TES chip to the bath. 
Special care has been taken for the design of the electrical circuit to provide a doubled up fault tolerant connection with low inductance microstrip traces on custom polymide PCBs (Fig.~\ref{fig:module} gray). 
The traces are further tinned with a Pb/Sn solder to provide a superconducting connection all the way from the 10\,mK feedthrough to the detector. 
The connections through the 10\,mK radiation shield are offset allowing a fill-in with Eccosorb\texttrademark~CR-110 to further minimize the leakage of IR radiation to the experimental volume. The remaining set of three connectors between the SQUID board and the detectors results in a typical parasitic resistance around 10\,m$\mathrm{\Omega}$ compared to a normal resistance of our sensors of a few 100\,m$\mathrm{\Omega}$ . 
 \begin{figure*}[htbp]
  \centering
  \includegraphics[height=0.37\textwidth]{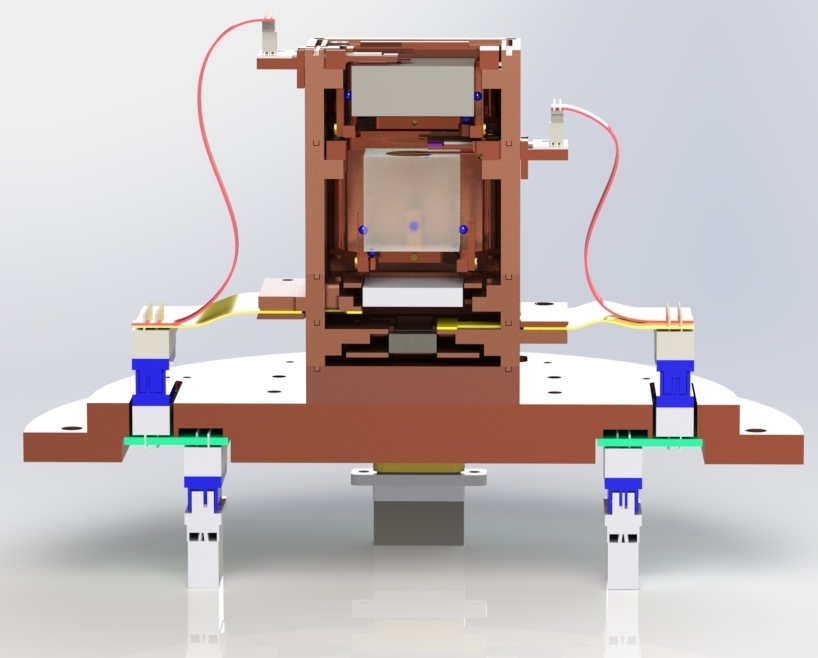}
  \includegraphics[height=0.37\textwidth]{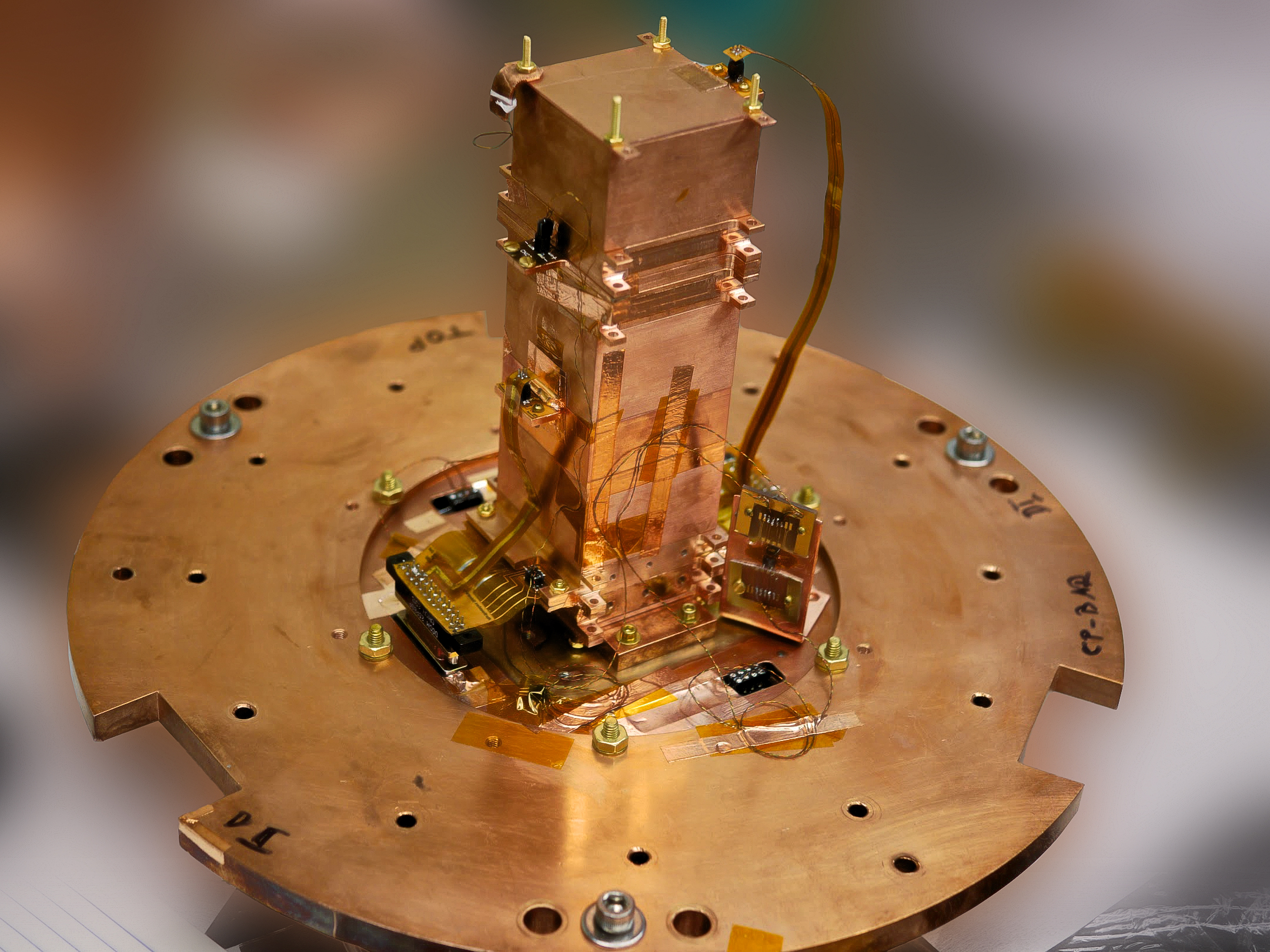}
  \caption{False color rendering of the lower part of the detector assembly showing the line of sight between the semi-transparent \lmo crystal and the silicon HVeV (white) detector below (left). Additional detectors on top and bottom pertain to SuperCDMS and Ricochet R\&D and are not discussed as part of this article. Picture (right) of the full detector assembly including auxiliary devices like 4-wire measurement samples and surplus thermometry . }
  \label{fig:design}
\end{figure*}

The detector package is stacked into a flexible height tower structure (Fig.~\ref{fig:design}), where an outer housing layer acts as main connection to the thermal bath and provides additional infra-red shielding. It is designed to be as light tight as possible to avoid a detrimental impact in the operation of Neganov-Trofimov-Luke (NTL) \cite{Neganov:1985, Luke:1988} assisted calorimeters from residual IR-photon background from the higher temperature stages of the cryostat \cite{Novati:2019}. Within a tower however, the design  allows for line of sight to the module below, where we placed an NTL-assisted $2 \times 2 \times 0.4$\,cm$^3$ Si HVeV (High Voltage and eV resolution) detector from the SuperCDMS collaboration (see \cite{Ren:2021}) to detect the scintillation light from the \lmo crystal.

\subsection{Light detector}
\label{sub:LightDetector}

In addition to the results from the operation of the \lmo detector with a TES sensor, we perform the first measurement of the scintillation time constants for this  \lmo candidate crystal for CUPID. 

\begin{figure}
\centering
\includegraphics[width=0.45\textwidth]{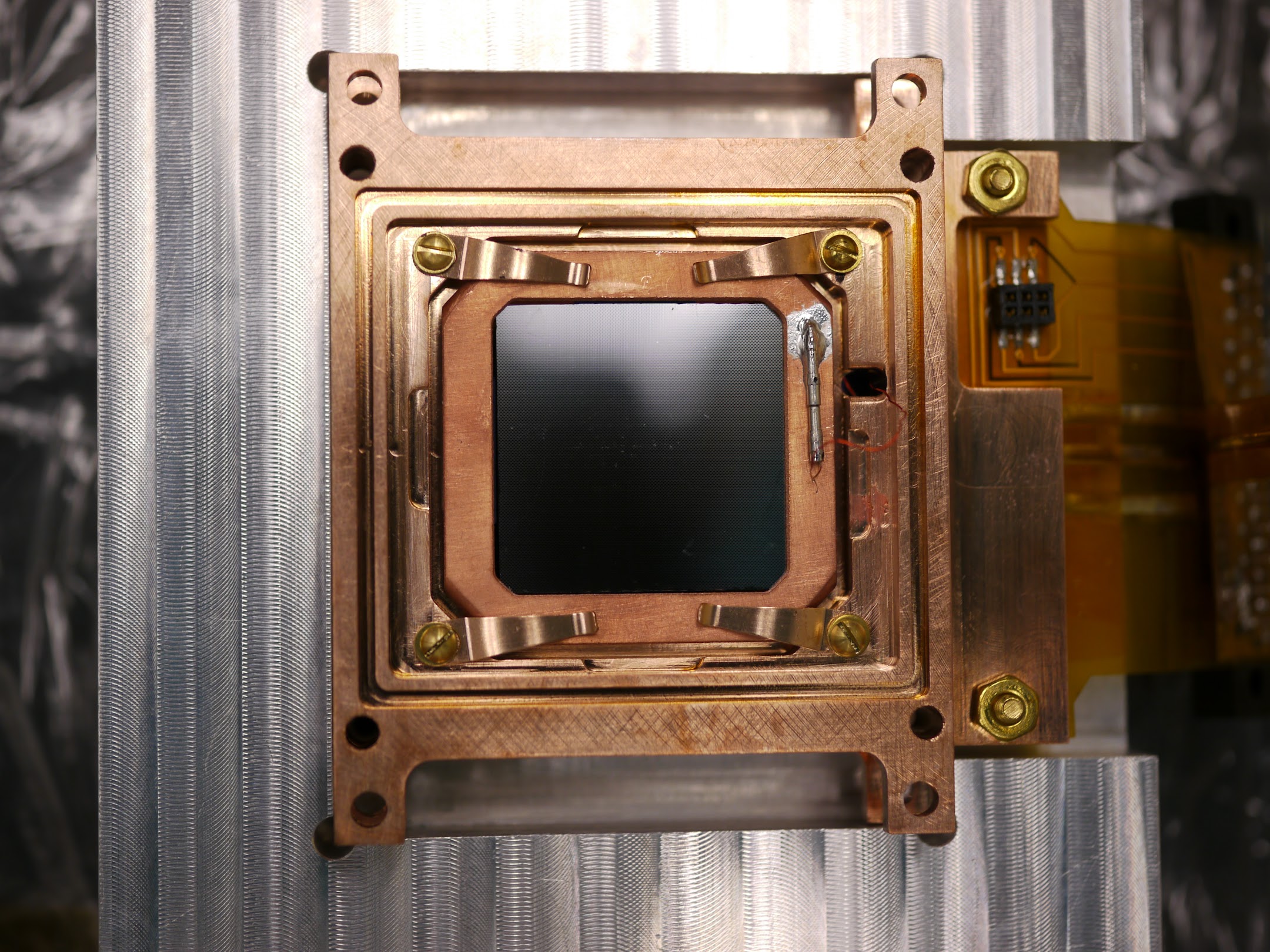}
\caption{Photograph of the 2\,cm side HVeV detector in its mounting jig. The image shows the back-side of the detector which faces the \lmo detector. A thin aluminium grid deposited on this side allows to apply a voltage bias across the detector.}
\label{fig:HVeV}
\end{figure}

The $2 \times 2 \times 0.4$\,cm$^3$ silicon HVeV detector shown in Fig.~\ref{fig:HVeV} is a larger version of the detector characterised in \cite{Ren:2021} and used for a dark matter search \cite{Albakry:2022} and a measurement of the nuclear recoil ionization yield in silicon down to 100\,eV nuclear recoils in \cite{Albakry:2023}. It allows for the application of a Neganof-Trofimov-Luke (NTL) \cite{Neganov:1985, Luke:1988} voltage bias across the device to convert and amplify an initial ionization signal into a larger heat signal. Furthermore, unlike the TES chip on our \lmo absorber, which was designed to primarily sense the thermal phonon component, these detectors use a parallel array of Quasiparticle trap assisted Electro-thermal-feedback Transition-edge sensors (QETs) \cite{Irwin:1995} developed for use within the SuperCDMS collaboration \cite{Ren:2021}. 
The device hence provides a significantly faster response with exponential rise- and decay-time constants of $\tau_r \approx 40$\,$\upmu$s, and $\tau_d \approx 250$\,$\upmu$s and it boasts an excellent signal to noise ratio of O(10) for the detection of individual photons when operated at 50\,V or larger NTL bias.
These characteristics make it very suitable for this measurement, despite its high reflectivity and a rather low photon collection efficiency in this setup. While these features would also be tremendously useful for CUPID itself, we note that these superb detectors are inherently complex in their fabrication and non-trivial to mass produce. We hence do not consider it as a first choice candidate for application in CUPID-1T and refrain from investigating the performance of the presented detector module in terms of heat-light characteristics and particle identification.

%-----------------------------------------------
\subsection{NEXUS low background cryogenic facility}
\label{sub:NEXUS}
The fully assembled detectors were installed at the Northwestern EXperimental Underground site in a Pulse Tube (PT) cooled dilution refrigerator from Cryoconcept. 
The system is equipped with Cryoconcept's proprietary Ultra-Quiet Technology\texttrademark, which uses the $^3$He/$^4$He gas mixture to thermally couple the PT head to the cryostat stages, while allowing an effective mechanical separation \cite{Olivieri:2017}. This decoupling is completed by a set of custom steel frames independently supporting the PT and cryostat while leaving only a soft bellow as mechanical contact.
The cryostat is a twin system to the Cryogenic Underground Test Environment (CUTE) at SNOLAB  \cite{Rau:2020, Camus:2024}. It was built with similar customizations with a 10\,cm thick internal low radio-activity lead shield and extended radiopure CuC2 copper vessels designed to fit a SuperCDMS tower. NEXUS thus features a large payload space of 30\,cm diameter $\times$ 65\,cm height and allows to maintain a base temperature of $\sim$10\,mK. The facility is intended for detector characterization measurements and has been designed to host a DD neutron-generator as well as a SiPM based backing array to perform kinematic neutron scattering measurements. It has been equipped with an external lead castle of 10\,cm thickness which allows the suppression of the ambient gamma flux from $^{40}$K and the U- and Th-chains by close to 2 orders of magnitude. This lead castle is comprised of a fixed wall on one side of the cryostat and a movable part mounted on rails to allow access to the cryostat and to take data in ``open" and ``closed"-shield configuration. NEXUS is hosted by Fermilab in the access tunnel of the MINOS experimental hall, which has a rock overburden of 107\,m (225\,m.w.e~\cite{Adamson:2015}) reducing the muon flux by more than two orders of magnitude to a value of ($0.80\pm 0.04) $\,muons/m$^2$/s \cite{Garrison:2014}. 
In addition NEXUS features both an external (METGLAS\textregistered) and an internal (AMUNEAL A4K\texttrademark) magnetic shield to avoid freezing in magnetic flux in either the SQUID amplifiers for the TES readout, the TESs themselves, or in components of the superconducting transmon qubits \cite{Bratrud:2024} and kinetic inductance detectors \cite{temples:2024} also studied in the facility.

For the readout of the TES detectors the cryostat is equip-ped with one SuperCDMS Detector Control and Readout Card \cite{Wilson:2022} that combines biasing, amplification, triggering and digitization of up to 12 SuperCDMS phonon channels in a single board. 
For R\&D purposes the firmware of our readout card at NEXUS has been modified to read out the currently installed set of 8 NIST SQUIDs in a continuous mode at 156.25\,kHz. 
The wiring is provided through Tekdata 24-wire NbTi looms, thermalized at 4\,K and 800\,mK before connecting to the SQUIDs at $\sim$90\,mK. Three out of the eight channels were dedicated to the 2\,cm HVeV detector and one for the \lmo detector discussed in this article. The remaining four channels were used for two further Ricochet TES-chip based detectors and a 1\,cm HVeV detector. The biasing circuit  was equipped with 10\,m$\mathrm{\Omega}$ shunt resistors for the Ricochet TES chips with a normal resistance of $\sim175\, $m$\mathrm{\Omega}$ and 20\,m$\mathrm{\Omega}$ shunt resistances for the SuperCDMS HVeV detector with a typical normal resistance of $\sim300$\,m$\mathrm{\Omega}$.

\section{Detector operation and performance}
\label{sec:performance}

The cool-down with the payload described in this article lasted over two months from October 17$^\mathrm{th}$ to December 10$^\mathrm{th}$ 2022. 
Despite significant time devoted to the optimization of the vibrational noise environment, we note the presence of a remaining non-negligible excess noise component revealed through pulse-tube on/off studies. 
The results presented here were obtained in six days of high energy calibration of the \lmo detector after the noise investigations. The detector was operated at an optimized working point of 52\,m$\mathrm{\Omega}$ (30\% R$_n$), with a bias current of 66\,$\upmu$A, at a stabilized MC temperature of 10\,mK. 
In order to obtain sufficient statistics in the photo-peaks in the MeV region with this 21\,g detector we took data with open shield utilizing the gammas from the $^{232}$Th chain and $^{40}$K from the rock of the cavern in addition to a lower energy $^{133}$Ba source. We also present results from the study of the scintillation time-constants of the \lmo crystal using a one-day run
taken with 60\,V Neganov-Trofimov-Luke amplification on the HVeV detector and within a reasonable noise environment.

%------------------------------------------------------
\subsection{Data processing}
\label{sec:processing}

\begin{figure}[htbp]
\centering
\includegraphics[width=0.49\textwidth]{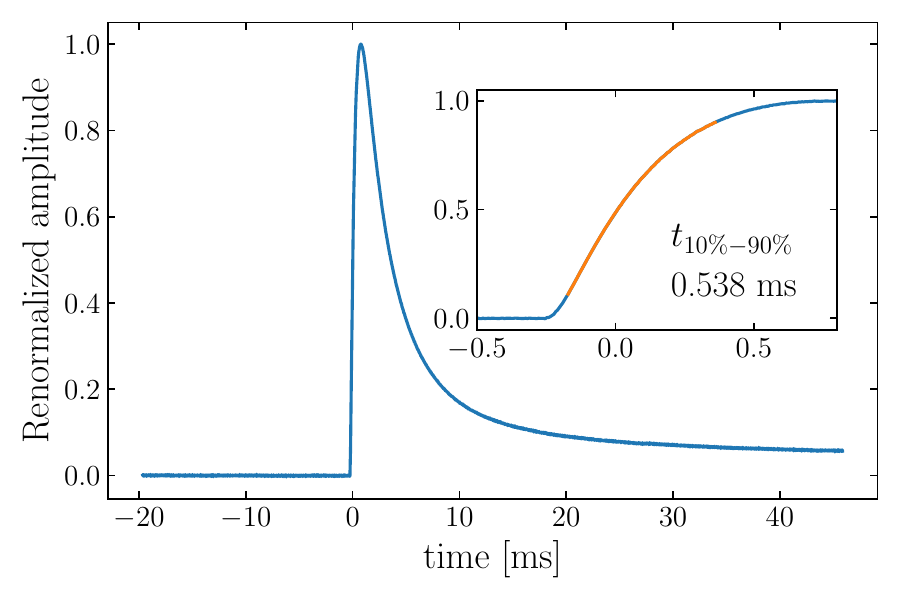}
\caption{Average pulse generated from events corresponding to the \K photo-absorption. The inset shows a zoom in of the first 1.25\,ms from the pulse, showing the sub ms rise-time. }
\label{fig:LMO_template}
\end{figure}

The data processing makes use of software tools and analysis steps similar to \cite{Ren:2021}. To trigger we utilize a robust Gaussian filtered derivative trigger optimized for the pulse-shape  of the \lmo detector with a gaussian filter time constant of 0.22\,ms. A pulse window of 19.7\,ms before trigger and 45.9\,ms after trigger provides a compromise between capturing the full return to baseline of the device and thus optimizing its energy resolution versus allowing us to run with a higher calibration rate to optimize statistics. 
The amplitude is estimated with an optimal filter algorithm \cite{Gatti:1986} with the average pulse selected from a region around the \K line in uncalibrated data. 
The detector response for these MeV-scale events shows a rise-time of $t_r$(10\% -- 90\%) = $538 \pm 11\,\upmu$s (Fig.~\ref{fig:LMO_template}). 
It was confirmed that the alignment and averaging provided no significant excess broadening based on the analysis of individual pulses.  
The noise power spectrum is built from a set of forced pseudo-random triggers rejecting remaining pulse like-events based on an iterative outlier removal as described in \cite{Ren:2021}. 

A live-time selection removes periods of time coincident with calibration measurements on the qubit payload and atypical noise periods. It was observed that resonance frequency scans mapping out the response of the RF-lines and connected qubits co-located in the refrigerator leads to a small cross-talk and pick-up on the TES readout lines. Removing these periods reduces the data by 24\% to 4.4 days.
To maximize statistics and assess the detector response up to the few MeV range, data was collected at a high calibration rate of 1.97\,Hz. 
Rejecting events with another trigger occurring within 45.9-ms before and after the triggered event, retains 78\% of the remaining data.

The high trigger rate combined with sub-optimal noise conditions of the cryogenic system with remaining excess vibrational noise induced by the pulse-tube required us to apply rather stringent analysis cuts:
three pulse shape parameters were considered, the pre-pulse mean baseline, its slope and the frequency domain $\chi^2$ of the pulse fit.
The selection on the mean of the baseline removes 10\% of data belonging to the upper tail of the  distribution. 
The selection on the slope keeps the events in the +/-3$\sigma$ region around the main population of the distribution. 
In order to ensure a correct amplitude estimation of the optimum filter, we also apply a pulse-shape based selection on the frequency-domain $\chi^2$.
This selection is energy dependent and removes the +3$\sigma$ tail of the $\chi^2$ distribution as shown in Fig.~\ref{fig:ChiSquareVsEnergy}. 
Combining these selections retains 67.8\% of all data, which is rather low compared to typical analysis efficiencies of cryogenic calorimeters of more than 90\%  for the search for \onbb~\cite{Adams:2022}. However, we expect to fully recover a similar efficiency in low rate data with improved cryogenic stability.
Given the sub-optimal run conditions, we accept a 6.8\% variation of the gain with respect to temperature changes after these cuts. Using the pre-pulse baseline as proxy for the detector temperature, we perform a linear correction of the change of gain based on the fit of this dependence at the \K line.
Similar to NTD-based detectors this linear gain stabilization \cite{Alessandrello:1998} is an important correction that significantly improves the energy resolution.  

\begin{figure}[htbp]
\centering
\includegraphics[width=0.49\textwidth]{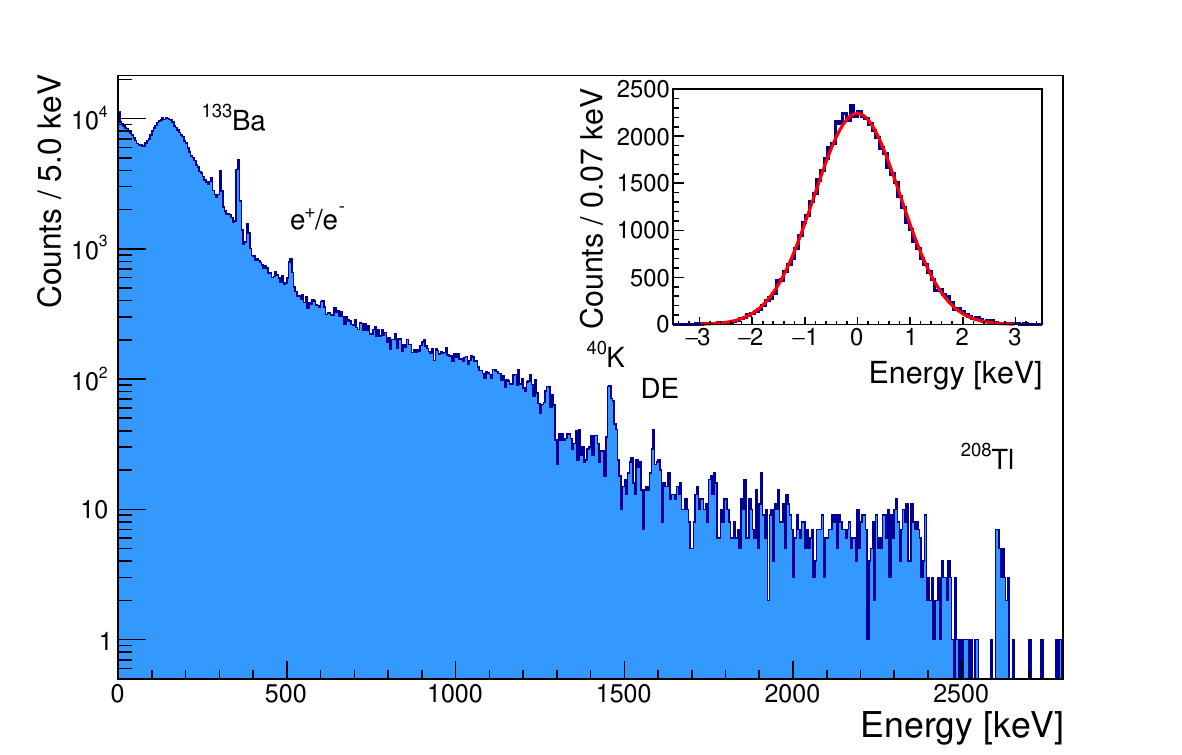}
\caption{Energy spectrum of 105\,hours of operation of the 2\,cm$^3$ \lmo detector with a $^{133}$Ba source and strong contributions of environmental $\upgamma$'s with open shield. The most intense calibration peaks are labeled: \TL provides the highest photopeak at 2.615 MeV and a Double Escape (DE) peak at 1.593 MeV.  The inset shows a fit of the baseline energy resolution estimated from forced pseudo-random triggers.}
\label{fig:LMOSpectrum}
\end{figure}

\begin{figure}[htbp]
\centering
\includegraphics[width=0.46\textwidth]{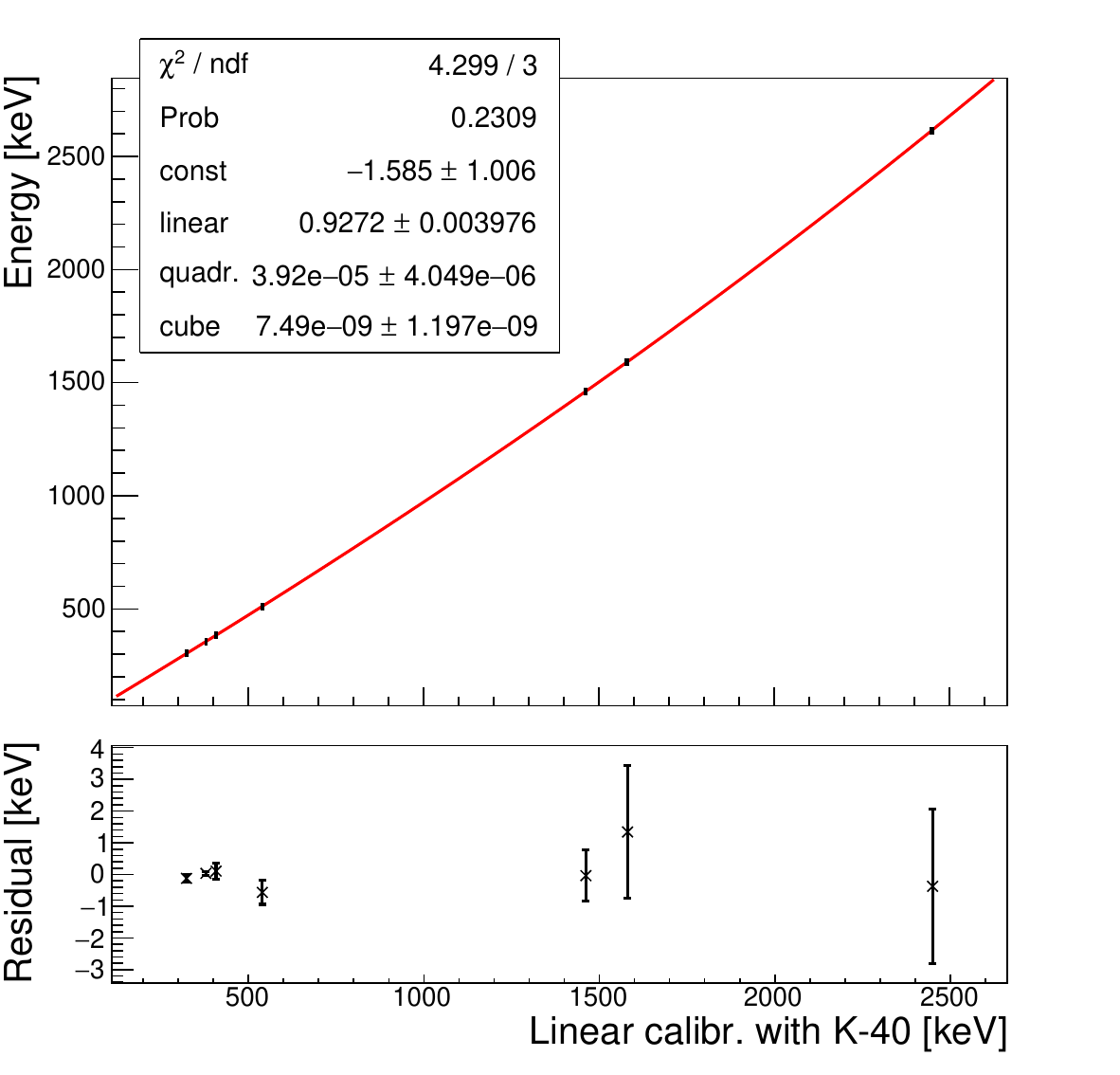}
\caption{Energy calibration and residuals of the calibration of of the 2 cm$^3$ \lmo detector with a third order polynomial.}
\label{fig:LMOCalibration}
\end{figure}

We calibrate this linearly-gain-stabilized spectrum using a third order polynomial to obtain the energy spectrum in Fig.~\ref{fig:LMOSpectrum}. The non-linear calibration --- reported in Fig.~\ref{fig:LMOCalibration} --- shows a good agreement with data, where the non-linear terms provide positive corrections at the level of 11\% (quadratic term) and 6\% (cubic term) for events at the Q value of \Mo (3034\,keV). We note that compared with a 4.5\,cm cubic CUPID crystal with $\sim$280\,g mass our 2\,cm crystal is expected to have less than 1/10$^{\mathrm{th}}$ of the heat capacity and covers a correspondingly larger resistance range of the TES's transition region in this measurement. The dynamic range of these TES chips is hence expected to be more than sufficient to cover both the $\upgamma$ and $\upalpha$ energy range for full size crystals.

%------------------------------------------------------
\subsection{Performance of the \lmo readout}
\label{sec:LMOperformance}

\begin{figure}[tbp]
\centering
\includegraphics[width=0.49\textwidth]{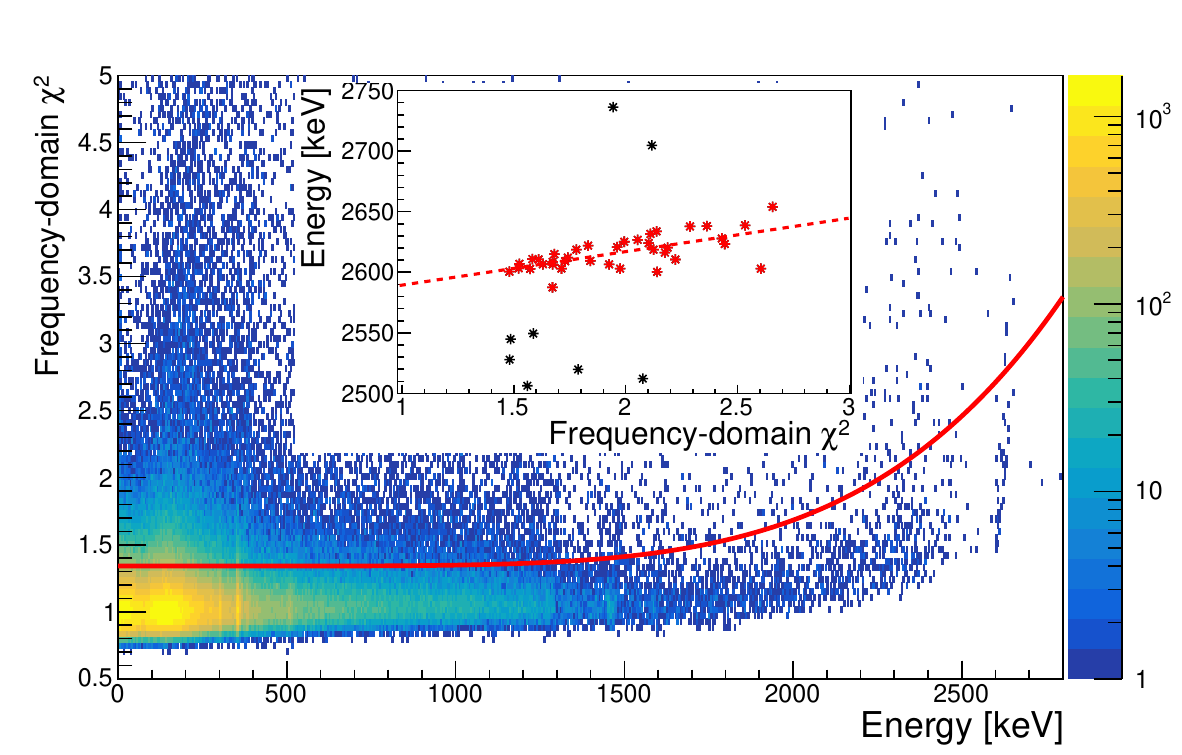}
\caption{Density plot of the reduced $\chi^2$ as quality of fit of individual events. Tails to high $\chi^2$ are caused by pile-up and thermal instabilities, while a mild worsening of the reduced $\chi^2$ above 1.5 to 2\,MeV suggests the onset of a non-linear detector response for this small 2\,cm crystal.}
\label{fig:ChiSquareVsEnergy}
\end{figure}

\begin{figure}[htbp]
\centering
\includegraphics[width=0.46\textwidth]{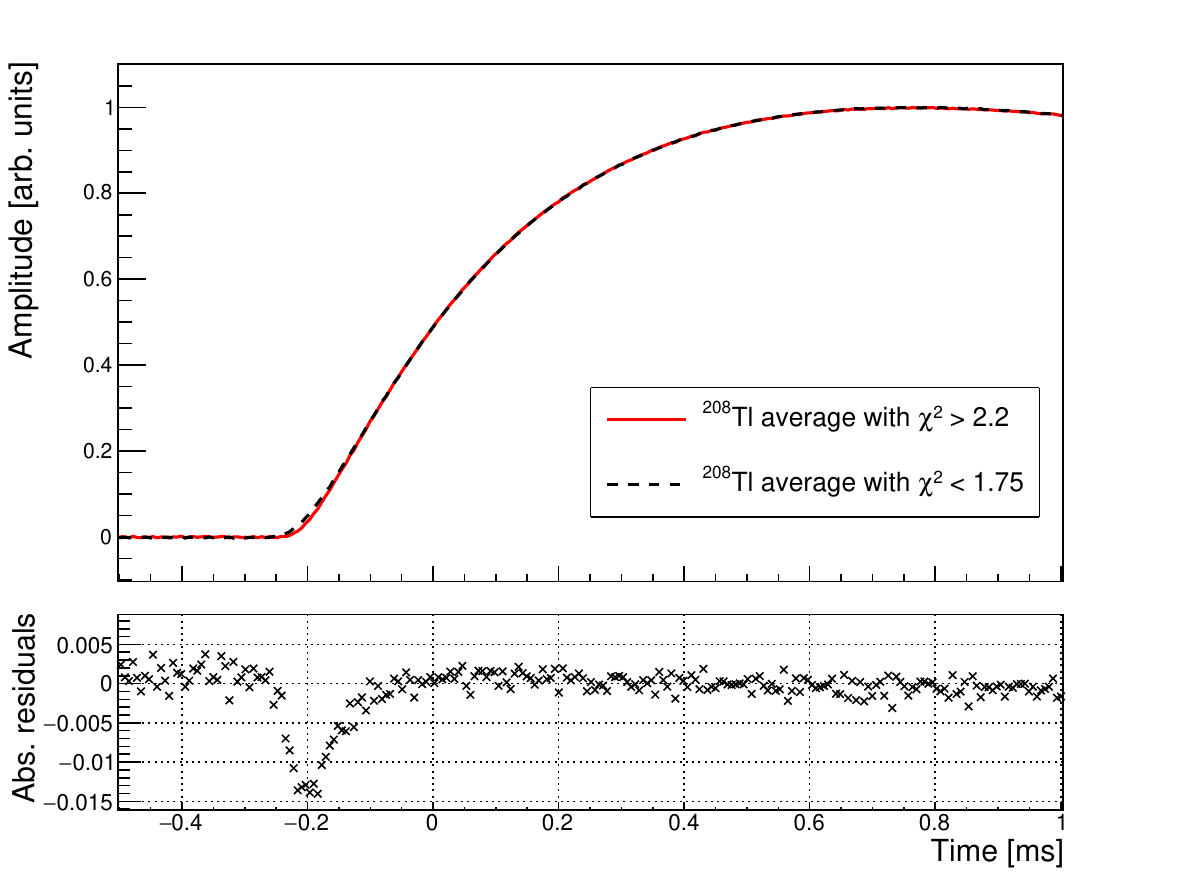}
\caption{Comparison and absolute difference between the normalized average pulse of events from the \TL line with $\chi^2 > 2.2 $ (red) and $\chi^2 < 1.75 $ (black). The residuals show the absolute difference of high-$\chi^2$ -- low-$\chi^2$. }
\label{fig:PluseShapeChi2Regions}
\end{figure}

We characterized the performance of the \lmo detector in terms of its resolution from the baseline noise up to the resolution at 2.615\,MeV and with respect to its pulse shape at high energy. 
We report the reduced $\chi^2$ as a function of energy in Fig.~\ref{fig:ChiSquareVsEnergy}. 
The flat band around a reduced $\chi^2=1$ represents the bulk of single events in stable conditions. All events below the red line are accepted for the further analysis. We note that the population of good particle events starts to exhibit an onset of energy dependence at around 1.8\,MeV. This is likely related to a mild onset of saturation or non-linear effects of the TES transition.
In addition, we observe a noticeable tilt of the high energy lines (2615\,keV, 1473\,keV) in  $\chi^2$ vs energy. This is highlighted in the inset in Fig.~\ref{fig:ChiSquareVsEnergy}, which shows a robust fit of the E vs $\chi^2$ dependence of the $^{208}$Tl photo-peak. 
We further investigate this effect creating an average pulse from two distinct regions in reduced $\chi^2$ below 1.75 (black) and above 2.2 (red) shown in Fig.~\ref{fig:PluseShapeChi2Regions}.   
As apparent from the absolute residuals of the two normalized averages $P_{\chi^2 > 1.8 }(t) - P_{\chi^2<1.75}(t)$ the distribution of events extending to higher $\chi^2$ not only provides an on average larger signal amplitude, but also a faster rise-time at the onset of the pulse. 
The red higher $\chi^2$ pulse-shape is aligned by its OF-maximum to the black pulse-shape. This causes it to show an apparent very slightly later onset which rises faster and quickly catches up to the normalized black pulse. 
This effect is limited to an initial period of $\sim200\,\upmu$s after which both shapes agree within uncertainty. We hypothesize that these slightly faster events are due to interactions in proximity of the gold film. These regions may be sensitive to a slightly larger ballistic phonon signal component compared to events farther from the gold film. We plan to further test this hypothesis in a future experiment with localized $\upalpha$-sources and multiple TES chips instrumenting a detector.

\begin{figure}[htbp]
\centering
\includegraphics[width=0.49\textwidth]{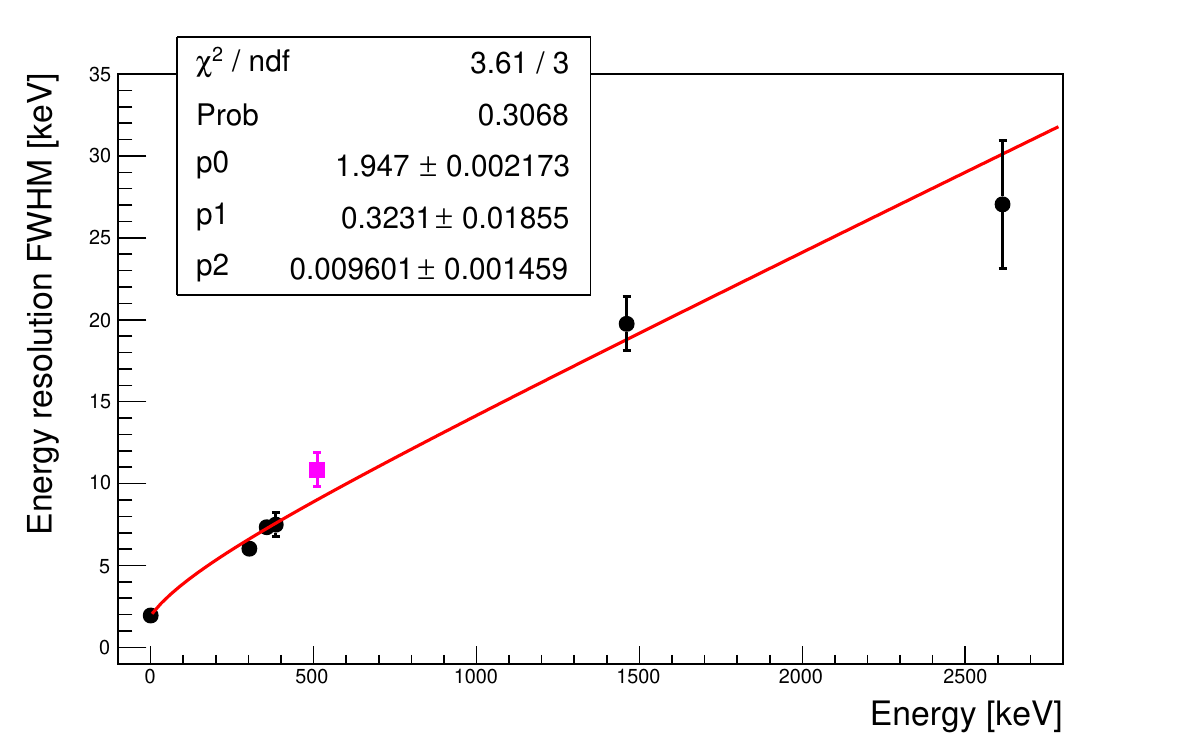}
\caption{Energy dependence of the measured energy resolution at the various calibration peaks fitted with $\sigma(E) = \sqrt{p_0^2 + (p_1\cdot \sqrt{E})^2 + (p_2\cdot E)^2}$. The 511-keV peak (magenta) from electron-positron annihilation has been excluded from the fit, to avoid any bias from Doppler broadening.}
\label{fig:LMOResolutionScaling}
\end{figure}

If left un-corrected this apparent pulse-shape difference results in an excess broadening of O(10\,keV) for the high energy peaks in the few MeV range. 
In Fig.~\ref{fig:LMOResolutionScaling} we show the resolution versus energy extracted from fitting a gaussian to the individual peaks at 2615\,keV $^{208}$Tl, 1461\,keV $^{40}$K, 384\,keV $^{133}$Ba, 356\,keV $^{133}$Ba, 303\,keV $^{133}$Ba, 276\,keV $^{133}$Ba and the baseline noise peak. 

The energy dependence is fit as:
\begin{equation}
    \sigma (E) = \sqrt{p_0^2 + (p_1 \cdot \sqrt{E})^2 + (p_2 \cdot E)^2},
\end{equation}
where the three terms can be interpreted as the sum of three independent noise contributions: a constant identified as the baseline resolution, a term scaling as $\sqrt{E}$, possibly connected to some statistical energy loss mechanism and a component scaling with $E$ which may be associated with position dependence. 
We note that the 511\,keV line has been excluded from the fit as it is expected that it may show some residual Doppler broadening. 
The baseline resolution is corrected for the reconstruction bias from the varying temperature reflected in the baseline slope variable and is measured as $1.947\pm0.002$\,keV (FWHM), similar to NTD based detectors operated in the past. The energy broadening 
scaling with $\sqrt{E}$ qualitatively reproduces the previously observed behaviour of \lmo crystals \cite{Armengaud:2021, Augier:2022}. The   linear dependence has a dominant contribution to the energy resolution above $\sim1500$\,keV. The fit with this additional contribution, not observed in high quality crystals \cite{Armengaud:2021} is preferred at more than $3\,\sigma$.

\begin{figure}[htbp]
\centering
\includegraphics[width=0.49\textwidth]{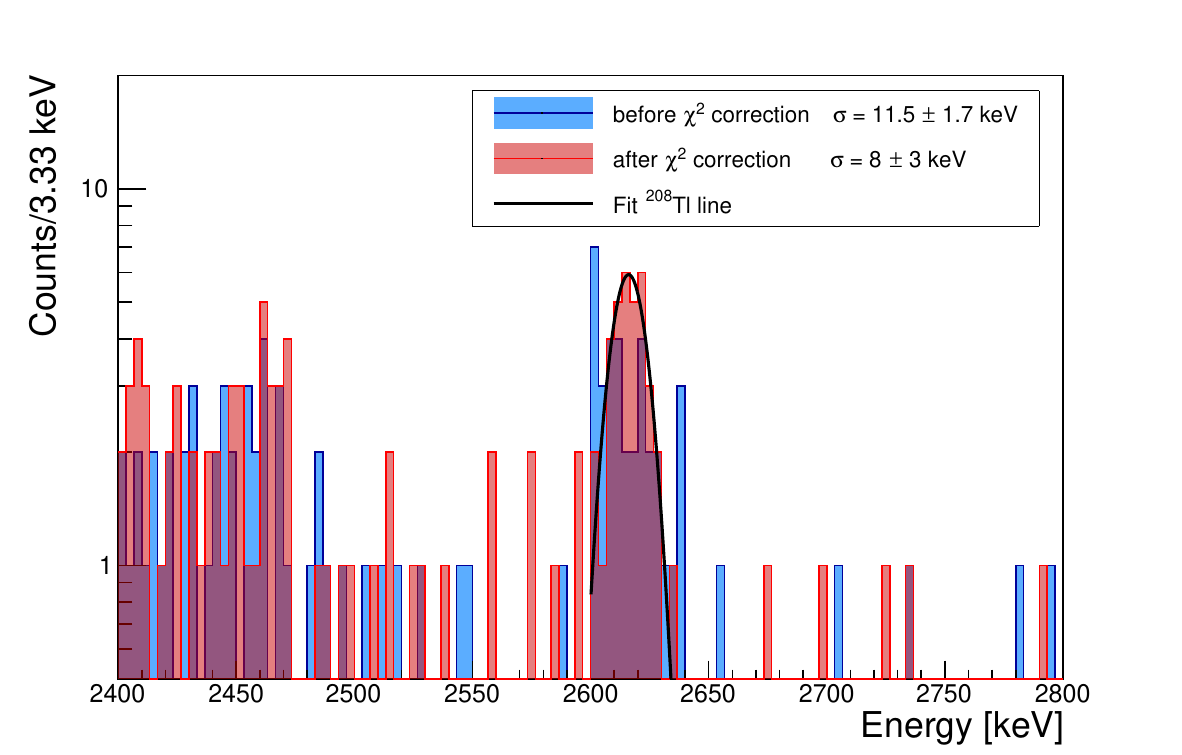}
\caption{Energy resolution of the $^{208}$Tl photo-peak before (blue) and after (red) correcting for pulse shape dependence.}
\label{fig:TlResolution}
\end{figure}

Utilizing the linear correlation observed between the $\chi^2$ and the energy of events in the \TL peak to introduce a pulse-amplitude correction we can improve the energy resolution at the \TL line by 30\% to a value of $\sigma = (8\pm3)$\,keV in the corrected spectrum shown in red in Fig.~\ref{fig:TlResolution}. It is tantalizing that this improvement matches the expected contribution of the linear component to the resolution broadening, but the statistical uncertainties are too large to draw any definite conclusions.
We note that the observed resolution broadening with $\sqrt{E}$, while qualitatively similar to the broadening observed for crystals grown by the Nikolaev Institute of Inorganic Chemistry (NIIC, Novosibirsk, Russia) for the CUPID-Mo experiment \cite{Schmidt:2020} is significantly more pronounced in the present test. It is unclear if this is related to the crystal itself, which is an early prototype out of the RMD crystal growth development program, or if it is related to the different readout. As will be discussed in Sec.~\ref{sec:ScintillationTimeConstants}, we do observe evidence for additional scintillation time constants compared to prior measurements, which could hint at the contribution of additional impurity centers to the light emission and which could also affect the bolometric performance.

%-------------------------------------------------------
\subsection{Rejection performance for \nnbb pile-up}
\label{sec:2nbbPileUp}

The detector architecture presented in this paper provides a significantly faster rise-time and decay-time compared to NTD-based detectors on the main absorber. In fact the rise-time is comparable or faster compared to NTD based light detectors, with a much better signal-noise ratio of O(3000) for the \onbb region of interest. To assess the potential impact on the expected background rate from random coincidence of the dominant \nnbb pile-up, we follow the same strategy as in \cite{Chernyak:2012,Chernyak:2017,Ahmine:2023}. We randomly sample a doublet of events from the \nnbb-spectrum which adds up to a total energy in the region of interest (around 3\,MeV) and inject them into the noise-data with a set of known fixed time separations. We evaluate the rejection efficiency for various time separations and find a better than 90\% rejection of pile-up events at 90\% single pulse acceptance for pulse separations as low as 45\,$\upmu$s. Assuming the same 10-90\% rise-time of $\sim$550\,$\upmu$s for a large mass (280\,g) CUPID-size crystal corresponds to an expected background index of $b=5 \cdot 10^{-6}$ counts/keV/kg/yr, an order of magnitude beyond the CUPID target of $b=5 \cdot 10^{-5}$\,counts/keV/kg/yr.

%----------------------------------------------------------
\subsection{\lmo scintillation time constants}
\label{sec:ScintillationTimeConstants}

\begin{figure*}[htbp]
\centering
\includegraphics[width=0.46\textwidth]{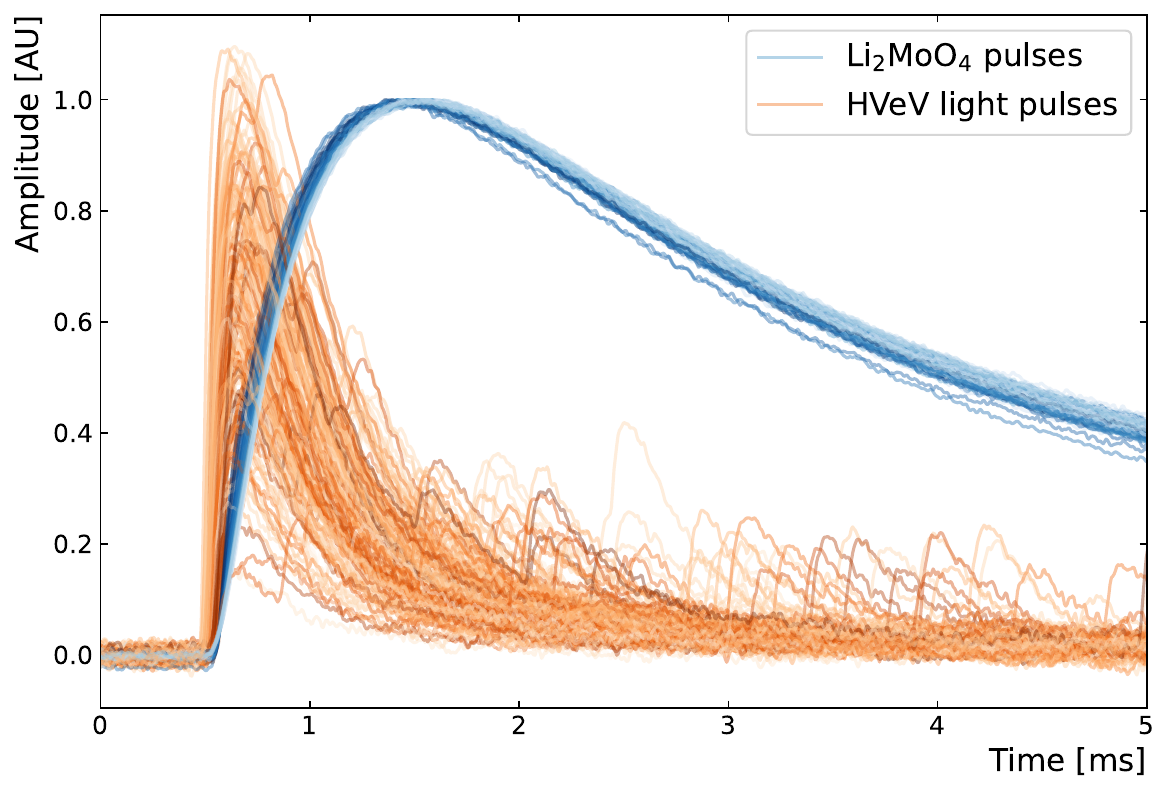}
\includegraphics[width=0.48\textwidth]{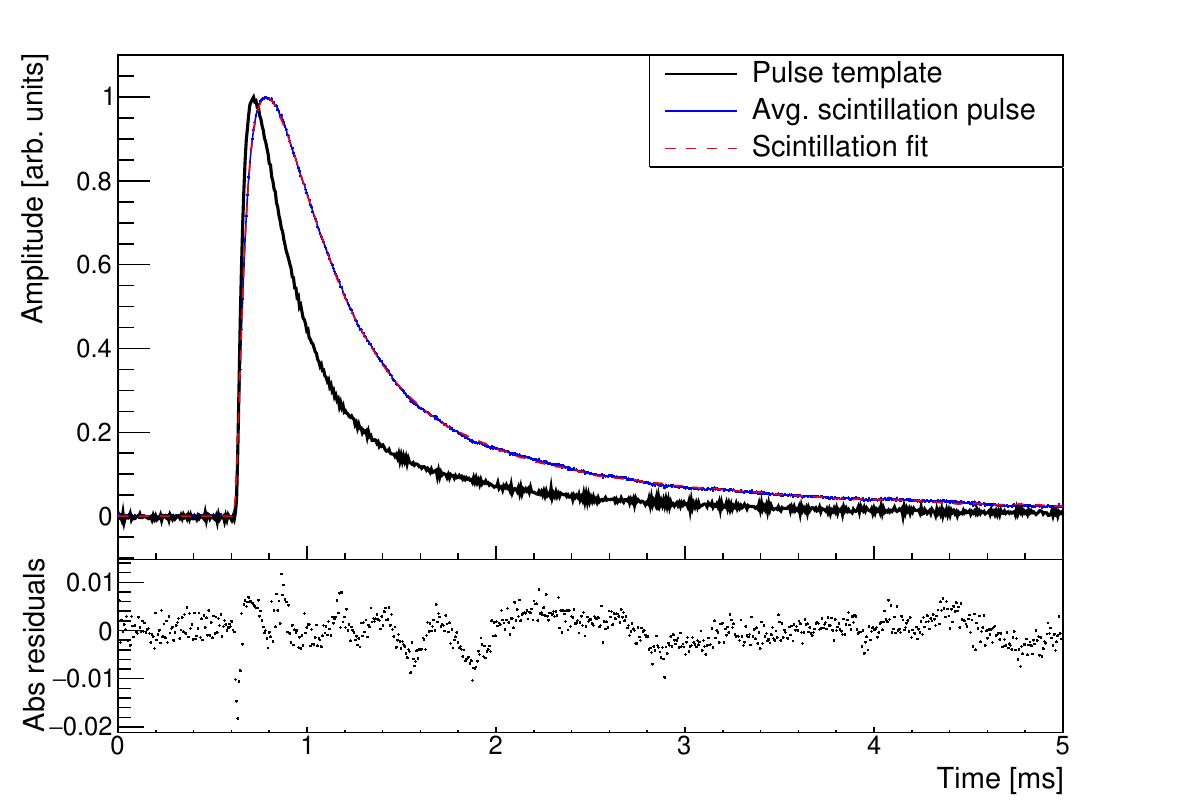}
\caption{Left: Coincident pulses of \lmo (blue) and scintillation light signals recorded with the HVeV detector (orange) with energy deposits within [900,1000] keV in the \lmo detector. The hue (bright to dark) reflects the start-time of the \lmo pulse. The \lmo pulses have been normalized to 1, while HVeV pulses have been scaled by 1/150.0 to preserve a consistent response to single photon events. The time resolution and signal to noise-ratio of the HVeV detector allows to see the arrival of individual photons from the scintillation decay constant on the tail of the pulse. Right top panel: Scaled pulse template from single electron-hole pair events in the HVeV light detector in anti-coincidence with the \lmo detector (black) and scaled average scintillation pulse from the [900-1000] keV region (blue). Fit of a triple exponential scintillation light yield model convoluted with the HVeV detector response (single electron-hole pair template) (red dashed); Right bottom panel: Absolute residuals between the averaged scintillation pulse and the fitted model.}
\label{fig:AvgScintillationFit}
\end{figure*}

In addition to assessing the detector performance for the novel TES based detector readout of the \lmo crystal, the present setup allowed to perform a precision measurement of the scintillation time constant of the RMD crystal.
We used two distinct measurement techniques for cross-validation: an average scintillation pulse fit and a photon trigger analysis.

\subsubsection{Average scintillation pulse fit}
\label{sec:AvgScintillationFit}
This method follows a similar approach as exploited with \edit{Metallic Magnetic Calorimeters} in \cite{Mailyan:2023} 
%with Kinetic Inductance Detectors in~\cite{Casali:2019} 
extracting the scintillation model from the measured scintillation pulse as a convolution of detector response and emitted scintillation intensity. 
\edit{This analysis relies on the fact that the light detector pulse shape (Fig.~\ref{fig:AvgScintillationFit} right in black) is well characterized, independent of particle type and position, and that the TES is operated within its linear regime. The consistent detector response has been ascertained for the SuperCDMS HVeV detector design through an extensive set of calibration measurements as described in \cite{Ren:2021, Albakry:2023,  Wilson:2024}. These detectors have also been used for several dark matter searches \cite{Albakry:2024, Amaral:2020, Agnese:2018} and as a benchmark scenario to test and improve Geant4 phonon simulation codes \cite{Kelsey:2023}. 
%The operation of the detector in NTL-mode, where the propagation of charge carriers and scattering provides the dominant phonon contribution, a homogenous detector response is guaranteed via the distributed phonon emission, but saturation effects can be observed c   
}
In the context of the data acquired in this work we use a numerical convolution of the measured HVeV detectors pulse template, extracted as average of single electron-hole pair events, with a multi-exponential scintillation model representing the scintillation light emission.
We report central values for the scintillation properties from the best fit of a selection of light detector and \lmo events with energies in the [900,1000]\,keV interval in Fig.~\ref{fig:AvgScintillationFit} (left). This interval provides the best compromise in terms of statistics, linearity of the HVeV detector response, signal-to noise and ease of alignment of the  pulses for their average. 
We observe clear evidence for multiple scintillation time constants fitting a scintillation model $S(t,t_0)$ with three exponential decay constants at a pulse start time $t_0$ using the Pymc Markov Chain Monte Carlo \cite{Pymc:2023} software package with
\begin{multline*}
S(t,t_0) = \Theta(t-t_0)\cdot \\\left( A_0\cdot \exp^{\frac{-(t-t_0)}{\tau_0}} +  
A_1\cdot \exp^{\frac{-(t-t_0)}{\tau_1}} + A_2\cdot \exp^{\frac{-(t-t_0)}{\tau_2}} \right)
\end{multline*}
Our reference fit is shown in Fig.~\ref{fig:AvgScintillationFit} (right). This fit uses our best \lmo pulse alignment procedure applying a butterworth low-pass filter to match up the \lmo pulses at 4\% of the pulse rise time and a \lmo pulse shape based statistical correction of the light detector start which corrects for a trend of the relative light detector pulse start time caused by the modified athermal \lmo pulse shape discussed earlier. Of particular importance for the search for \onbb is the fact that the rise-time of the scintillation pulse (34\,$\upmu$s, 30\% - 70\%) shows only a very minor broadening compared to the intrinsic detector response of our fast athermal phonon light detector (20\,$\upmu$s, 30\% - 70\%). It implies that for all presently considered light detectors for CUPID with rise-times at the level of several 100\,$\upmu$s or more, pile-up rejection based on scintillation light is entirely dominated by the bandwidth and response time of the detector.

\edit{Our fit results }in a fast scintillation time constant of \begin{equation*}\tau_0 = \left( 17.2 \pm 0.2~\mathrm{(stat.)} ^{+35}_{-14}~\mathrm{(syst.)}\right)\,\upmu \mathrm{s} \end{equation*} contributing $42^{+9}_{-7}$\% of the detected scintillation light.  
The slower scintillation constants in turn have characteristic time scales of 
\begin{equation*} \tau_1 = \left(203\pm 3~\mathrm{(stat.)} ^{+67}_{-43}~\mathrm{(syst.)}\right)\, \mathrm{\upmu s}
\end{equation*} and 
\begin{equation*} \tau_2 = \left(2.3\pm 0.09~\mathrm{(stat.)} ^{+1.1}_{-0.7}~\mathrm{(syst.)} \right)\,\mathrm{ms}. 
\end{equation*} 
contributing $46^{+6}_{-7}\%$ and $12^{+2}_{-3}\%$ respectively.
The reported systematic uncertainties have been determined from additional fits falling into two major categories: (a) fits of 5 different event samples from different energy slices from [500, 600] to [2000, 2600] keV using the same analysis methodology (b) fits of the [900, 1000] keV event sample with different analysis methodology. Results are given in Table~\ref{tab:Systematics}. 
\tabcolsep=0.18cm
\begin{table}
\centering
\caption{Reference fit result, and systematic uncertainties evaluated from fits of an average scintillation pulse out of 5 additional energy slices - Category (a) - and from changes in the analysis method with respect to the alignment in the computation of the average pulse - Category (b). Statistical uncertainties are negligible in comparison. f$_i$ are the fractions of the total scintillation light emitted by the fast, medium and slow contribution within the 5\,ms window. The combined  uncertainty (stat. and syst.) from addition in quadrature is shown in the final row.}

\begin{center}
\begin{tabular}{c|c|c|c|c|c|c}
\hline\noalign{\smallskip}
Category  & $\tau_0$  & $\tau_1$ & $\tau_2$ & f$_0$ & f$_1$  & f$_2$\\
 & ($\upmu$s) & ($\upmu$s) & (ms) & (\%) & (\%) & (\%)\\
\hline
Reference fit &  17 & 203 & 2.3 &  42 &  46 &  12  \\ \hline
Syst. (a) &  $^{+13}_{-12}$ & $^{+65}_{-42}$ & $^{+1.1}_{-0.7}$ &  $^{+8}_{-4}$ &  $^{+2}_{-6}$ &  $^{+2}_{-3}$  \\ 
Syst. (b) & $^{+32}_{-8}$ & $^{+16}_{-8}$ & $^{+0.3}_{-0.2}$  &  $^{+4}_{-6}$  &  $^{+6}_{-4}$  &  $^{+4}_{-5}$  \\ 
\hline
Combined & 17$^{+35}_{-14}$ & 203$^{+67}_{-43}$ & $2.3^{+1.1}_{-0.7}$  &  $42^{+9}_{-7}$  &  $46^{+6}_{-7}$  &  $12^{+2}_{-3}$ \\ \hline
\end{tabular}
 \label{tab:Systematics}
\end{center}
\end{table}

Category (a) accounts for systematics related to our event sample like baseline drifts or the possibility of unidentified particle coincidence events instead of a pure scintillation light event sample for the [900, 1000]\,keV event selection. It also encompasses a test of energy dependent effects or imperfections in detector response \edit{and finally, the leading systematic, the definition of the event start time through the \lmo pulse alignment.} We expect consecutively better alignment with higher signal to noise ratio for the higher energy event samples, but in turn lower overall statistics with only $\sim60$ events in our energy slices above 1.3 MeV and an expected onset of non-linear detector response effects in the HVeV detector.

Category (b) focuses on systematics from the analysis methodology, in particular the dominant contribution from the algorithm for aligning the pulses in the average scintillation pulse. 

We use the following two different analysis methods: (1) Aligning the \lmo pulses by the optimum filter algorithm and including the remaining jitter of the pulse start time as a gaussian smearing in the fit model. We estimate this residual jitter from the low-pass filtered \lmo pulses at 4\% and 30\% of the pulse height. (2) Aligning the events with the low-pass filtered pulse at 4\% of the pulse rise, but foregoing the pulse-shape based correction that accounts for athermal pulse shape dependent offsets in the light detector start time with respect to the measured \lmo start time.
We estimate the uncertainty from the fit results in the first category assuming that the results with positive (negative) deviations follow independent half-normal distributions and extract the systematic uncertainty as $\sigma^{+/-} = \frac{\sqrt{\pi}}{\sqrt{2}} \cdot \overline{\tau_i^{+/-}}$, where $\overline{\tau_i^{+/-}}$ is the mean of all fit results with positive (negative) deviations when compared to the reference fit. For the second category of correlated or identical data we conservatively use the maximum positive or negative deviation and combine the uncertainty of both categories in quadrature. 
We note that our results are restricted to the 5\,ms window which contains the bulk of the scintillation light emission and is of primary interest for \onbb applications. 
Coincidence events and drifts of the baseline make the analysis of longer time-scales \edit{subject to increased systematic uncertainties}.
We further investigated the sensitivity of our MCMC fit to varying initial parameters and varying sets of priors, observing no significant variation.

\subsubsection{Photon trigger analysis}

\begin{figure}[htbp]
\centering
\includegraphics[width=0.48\textwidth]{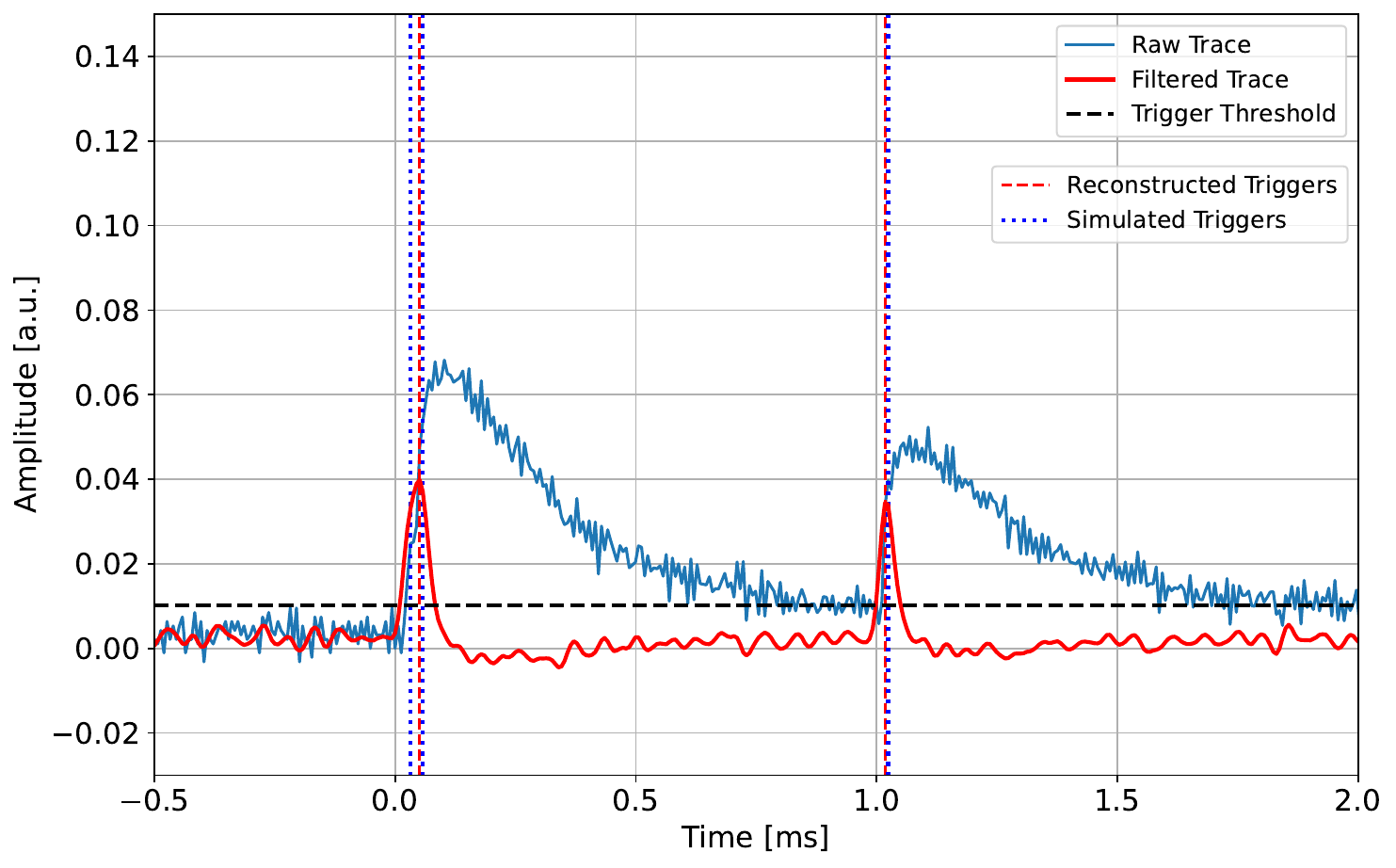}
\includegraphics[width=0.48\textwidth]{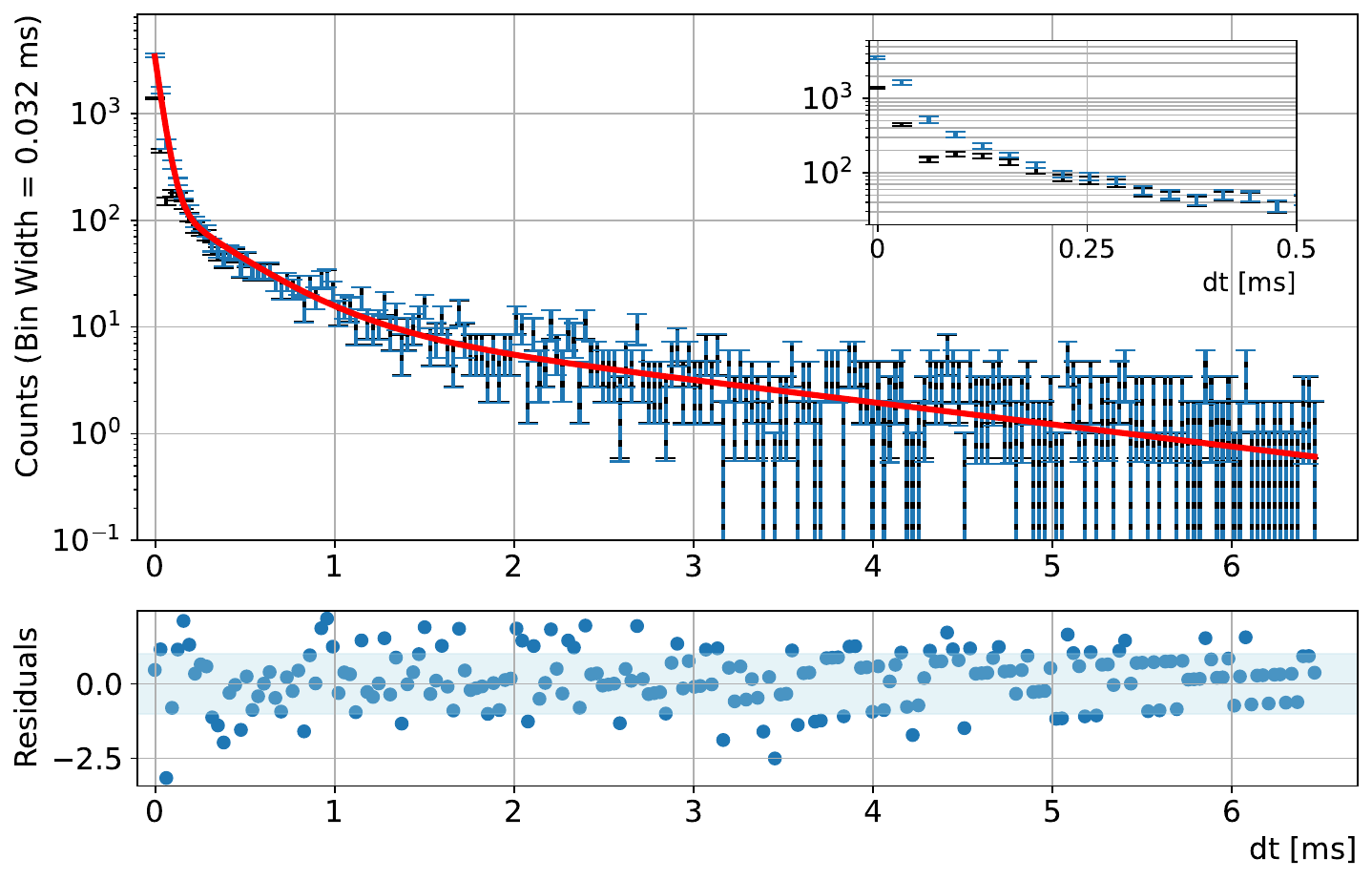}
\caption{  \edit{Top: Example of a synthetic pulse simulation to statistically estimate the trigger inefficiency for photons arriving in quick succession. From the three simulated triggers highlighted by the blue dotted lines only two are reconstructed as individual distinct triggers in the gaussian derivative filtered trace (red). \\
Bottom:  Distribution of trigger times of detected photons from the \lmo scintillation light with respect to the \lmo event time (black). Efficiency corrected photon arrival time distribution in (blue) with the impact on the initial reconstructed triggers in the first few hundred microsecond highlighed in the inset. The fit (red) is a triple exponential scintillation model with residuals in standard deviations on the bottom. The blue band corresponds to the $\pm1\,\sigma$ range. } }
\label{fig:ScintillationConstants}
\end{figure}

The second technique makes use of the excellent signal to noise ratio of the HVeV detector operated at 60\,V NTL amplification to analyze the arrival and trigger time distributions of individual photons (Fig.~\ref{fig:ScintillationConstants} bottom) with a very similar procedure to \cite{Albakry:2022}. The baseline resolution of 2.3\,eV phonon resolution before NTL amplification  together with the work carried out by the drift of the charge carriers in the electric field results in an energy deposit of $\sim$62\,eV per photon and in turn in a fractional charge resolution of $\sim$10\% at the single electron-hole pair peak. Expressed as an equivalent resolution it would correspond to a resolution of 0.2\,eV for the average photon energy of 2.1\,eV for \lmo scintillation \cite{Bekker:2016}. Individual photons are hence clearly identifiable in the \edit{low photon number regime of the scintillation light pulses.
However, the fast scintillation component results in a non-negligible amount of photons arriving in very quick succession in the first several tens of $\upmu$s of a scintillation event. The resulting HVeV pulses overlap in their rise and are difficult to trigger individually as shown with a pulse simulation in Fig.~\ref{fig:ScintillationConstants} (top). 
To limit the number of pile-up events we hence 
 chose to fit events from the [500,700]\,keV region with lower average number of photons incident on the HVeV detector ($\sim 3.9$) per event. 
In addition we include an efficiency correction based on an iterative procedure of simulating the scintillation data and finding a range of scintillation parameters which reproduce the observed data after applying the trigger algorithm (see  \cite{Albakry:2022}, for details on the procedure).}
The effect of this correction \edit{is highlighted in the inset} in Fig.~\ref{fig:ScintillationConstants} bottom in blue. 
The triple exponential scintillation model fit over a range of 6.5\,ms is consistent with the results from our primary method of fitting the average scintillation pulse. It results in a measurement of 
\begin{align*}
    \tau_0 &= \left(36\pm  2~\mathrm{(stat.)} ^{+14}_{-10}~\mathrm{(syst.)}\right)\, \mathrm{\upmu s} \\
     \tau_1 &= \left(340\pm  38~\mathrm{(stat.)} ^{+149}_{-97}~\mathrm{(syst.)}\right)\, \mathrm{\upmu s} \\
      \tau_2 &= \left(2.1\pm  0.3~\mathrm{(stat.)} ^{+0.9}_{-0.4}~\mathrm{(syst.)}\right)\, \mathrm{ms}. 
\end{align*}
The systematics associated with this method are complementary to our average scintillation pulse fit. 
The photon arrival time fit is nearly unaffected by baseline instabilities which can bias the estimation of the long scintillation time constant in our first method. 
\edit{However, the trigger efficiency correction results in a correction of the initial bins in the photon arrival fit by a factor 2 to 5 and puts a significant uncertainty on the amplitude and relative contribution of the first scintillation time constant. Through anti-correlations, this also affects the relative contributions of the slower time constants.} The estimated contributions to the overall light yield from $\tau_0$, $\tau_1$ and $\tau_2$  are $f_0 = 59^{+13}_{-10} \%$, $f_1 = 26^{+8}_{-9} \%$, and  $f_2 = 15 \pm 7 \%$. Individual systematics tests of the methodology in this photon arrival time fit are listed in Table~\ref{tab:Systematics2}. 
\tabcolsep=0.18cm
\begin{table}
\centering
\caption{Reference fit result with statistical uncertainty and systematic uncertainties evaluated for the trigger reconstruction efficiency $\upepsilon$ (Reference, low $\upepsilon$, high $\upepsilon$ ) and changes in the analysis method varying the start of the fit range by up to ($\pm 12.8\,\upmu$s) and the remaining jitter between light detector and \lmo pulse by ($12\pm 6.4\,\upmu$s). We also list the combined (stat. and syst.) uncertainty in the final row.}
\begin{center}
\begin{tabular}{c|c|c|c|c|c|c}
\hline\noalign{\smallskip}
Type  & $\tau_0$  & $\tau_1$ & $\tau_2$ & f$_0$ & f$_1$  & f$_2$\\
 & ($\upmu$s) & ($\upmu$s) & (ms) & (\%) & (\%) & (\%)\\
\hline
Ref.~$\upepsilon$ fit &  36$\pm2$ & 340$\pm38$ & 2.1$\pm0.3$&  59$\pm3$ &  26$\pm6$ &  15$\pm3$ \\ 
\hline
Fit start &  $^{+11}_{-4}$ & $^{+91}_{-38}$ & $^{+0.5}_{-0.1}$ &  $^{+1}_{-1}$ &  $^{+2}_{-0}$ &  $^{+0}_{-1}$  \\ 
Jitter & $^{+0}_{-6}$ & $^{+1}_{-59}$ & $^{+0.0}_{-0.3}$  &  $^{+0}_{-2}$  &  $^{+1}_{-0}$  &  $^{+2}_{-0}$  \\ 
\hline
Low $\upepsilon$ fit &  42$\pm2$ & 434$\pm55$ & 2.7$\pm0.6$&  70$\pm4$ &  20$\pm5$ &  10$\pm2$ \\ 
\hline
Fit start &  $^{+8}_{-1}$ & $^{+55}_{-21}$ & $^{+0.4}_{-0.0}$ &  $^{+1}_{-3}$ &  $^{+2}_{-1}$ &  $^{+1}_{-1}$  \\ 
Jitter & $^{+0}_{-3}$ & $^{+5}_{-28}$ & $^{+0.0}_{-0.2}$  &  $^{+1}_{-0}$  &  $^{+0}_{-1}$  &  $^{+0}_{-0}$  \\ 
\hline
High $\upepsilon$ fit &  33$\pm2$ & 300$\pm37$ & 1.9$\pm0.2$&  52$\pm3$ &  30$\pm6$ &  18$\pm3$ \\ 
\hline
Fit start &  $^{+12}_{-5}$ & $^{+95}_{-48}$ & $^{+0.5}_{-0.2}$ &  $^{+2}_{-1}$ &  $^{+1}_{-0}$ &  $^{+1}_{-2}$  \\ 
Jitter & $^{+0}_{-6}$ & $^{+4}_{-57}$ & $^{+0.0}_{-0.2}$  &  $^{+0}_{-4}$  &  $^{+0}_{-0}$  &  $^{+2}_{-0}$  \\ 
\hline
Combined & $36^{+14}_{-10}$ & $340^{+154}_{-104}$ & $2.1^{+0.9}_{-0.5}$  &  59$^{+13}_{-10}$  &  $26^{+8}_{-9}$  &  $15^{+7}_{-7}$  \\ 
\hline
\end{tabular}
 \label{tab:Systematics2}
\end{center}
\end{table} 
The overall systematic uncertainty is conservatively estimated as largest positive/negative deviation from all systematic tests. Overall, the two methods are in good agreement confirming a multi-exponential scintillation model with a significant fast initial light emission.

\section{Discussion \& Outlook}
\label{sec:outlook}

The detector performance discussed in the present article demonstrates several crucial aspects for a potential application of this detector in \onbb search:  
the TES-based 2\,cm \lmo detector achieved a promising baseline resolution ($1.947 \pm 0.002$\, keV FWHM) for a first test, compatible with the resolution of the NTD based CUPID-Mo detectors \cite{Armengaud:2020}. 
The response of the TES sensor also showed a reasonable non-linearity and a dynamic range that can easily cover the MeV range, even with a small 21\,g detector. 
The detector displayed a more significant excess broadening with energy though when compared to previous \lmo detectors from NIIC, but similar differences have been observed in other detector prototypes with NTD based detectors. 
Both the surplus scintillation time constants and the pulse-shape of the phonon signal of the \lmo detector itself, which does not return to the baseline within the pulse-window suggests additional impurities that could degrade the detector response. A future test with a reference crystal from NIIC, that has previously been operated with NTD's is hence planned to resolve the origin of this resolution broadening. Furthermore, several design improvements of the TES chip are being considered.

We observed evidence for position dependence in the energy response and demonstrated that a preliminary correction based on the frequency domain $\chi^2$ can be used to correct for it. 
We plan to perform more detailed studies of this effect to evaluate different pulse shape based variables, like a PCA based analysis of $\sim$100 samples around the onset of the pulses. 
As demonstrated in \cite{Huang:2021} such a method can be used to identify and quantify the most significant shape deviations and can consequently be used for efficient corrections or data quality cuts. 
We plan to use these methods and take new data with localized-surface and extended-bulk event sources to confirm our hypothesis that this position dependence is due to the collection of an athermal phonon component for events in close proximity of the gold collector film. 
In addition we plan to study the extent of this effect in terms of the fiducial volume fractions of events strongly/barely affected by this athermal contribution, possibly through the use of multiple sensors on the same detector.  
We note that the possibility of position and event topology reconstruction with cryogenic calorimeters with multiple sensors has been suggested in \cite{Armatol:2022}.

Finally, the present design boasts a significantly faster readout both in rise- and decay-time thanks to the electro-thermal feedback of the TES detector and we note that it is inherently tunable in terms of speed and optimal resolution~\cite{Chen:2023}.
Since this TES carrier chip based readout is also applicable to the instrumentation of a silicon or germanium-wafer based light detector, we plan to explore an optimization for the best resolution in the \lmo and minimal rise-time for a Neganov-Trofimov-Luke amplified light detector in a future set of measurements. The observed \lmo scintillation emission showed a fast pulse rise-time (30-70)\% that was only 14\,$\upmu$s longer than the intrinsic detector response of our light detector, much below the rise-time of currently considered light detectors for CUPID.  \edit{The presence of a measured fast scintillation constant of O(20\,$\upmu$s) agrees with reported values in \cite{Mailyan:2023, Chen:2018, Spassky:2015}.  However, there is disagreement on the relative contribution and time-scale of a slower scintillation component \cite{Mailyan:2023, Spassky:2015} if observerd at all \cite{Chen:2018, Casali:2019}. Spassky et al. \cite{Spassky:2015} noted distinct differences in the scintillation properties for  \lmo{} samples grown via Low Thermal Gradient Czochralski vs Czochralski samples. 
%is also about four times faster than previously reported values for \lmo crystals from NIIC in \cite{Casali:2019}. 
A follow up measurement with consistent instrumentation and methodology and \lmo detectors from different growth procedures would be valuable to ascertain if the reported differences are due to the difference in crystal growth or related to the methodology of the analysis. }
If confirmed the fast scintillation times would allow for significant further improvements on the pile-up rejection with better light detectors.
This technology thus has the potential to provide the basis for a full TES-based design of a CUPID-1T size experiment. 
It could enable an experiment with a single, scalable, multiplexed readout solution, with the required background levels at O$(10^{-6}$\,counts/keV/kg/yr) with respect to \Mo pile-up.

\begin{acknowledgements}
This work is supported in part by NSF grant PHY-2209585, the Connaught Fund at University of Toronto, and the Canada First Research Excellence Fund through the Arthur B. McDonald Canadian Astroparticle Physics Research Institute.  This research was enabled in part by support provided by SciNet (www.scinethpc.ca) and the Digital Research Alliance of Canada (alliancecan.ca). Funding and support were received from NSERC Canada. This work is a synergetic development with R\&D for the RICCOCHET experiment. 
The work made use of the NUFAB facility of Northwestern University’s NUANCE Center, which has received support from the SHyNE Resource (NSF ECCS-2025633), the IIN, and Northwestern’s MRSEC program (NSF DMR-1720139).
Work at Argonne National Lab, including work performed at the Center for Nanoscale Materials, a U.S. Department of Energy Office of Science User Facility, is supported by the U.S. Department of Energy, Office of Science, Office of High Energy Physics and Office of Basic Energy Sciences, under Contract No. DE-AC02-06CH11357. Work at MIT is supported by the U.S. Department of Energy under Contract No. DE-SC0011091.

We thank RMD Inc. for the production of the crystal and discussions on the growth of \lmo.
We  thank Dr.~Anastasiia Zolotarova and the NUFAB staff for initial discussions that helped the authors to implement the gold deposition procedure on \lmo. Finally, the authors want to gratefully thank Dr.~Dan Bauer, for his role in conceiving and commissioning the NEXUS facility together with Prof.~Enectali Figueroa-Feliciano and his continued advice and support out of his retirement during NEXUS operations.

\end{acknowledgements}

\bibliographystyle{spphys}       
\bibliography{Bibliography.bib}

%\newpage
%\include{AnswersToReferee}

\end{document}